\documentclass[aps,prd,floatfix,showpacs,10pt,nofootinbib]{revtex4}
\usepackage{epsfig}
\usepackage{amsmath,amssymb,amsfonts,amsthm} 
\usepackage{color}
\usepackage{float}
\usepackage{graphicx}
\usepackage{pst-plot}
\usepackage[scriptsize,nooneline,hang]{caption}
\usepackage[hang,nooneline,scriptsize]{subfigure}

\definecolor{dark-green}{rgb}{0,0.7,0}
\definecolor{dark-blue}{rgb}{0,0.2,0.5}
\definecolor{med-blue}{rgb}{0,0.7,1}
\definecolor{mblue}{rgb}{0,0.2,1}
\definecolor{cnc}{rgb}{0.8,0,0}
\definecolor{light-red}{rgb}{1,0.8,0.8}
\definecolor{dark-yellow}{rgb}{1,0.8,0}
\definecolor{light-blue}{rgb}{0.8,0.9,1}
\definecolor{grey}{rgb}{0.211,0.211,0.211}
\definecolor{verylight-blue}{rgb}{0.93,0.95,1}
\definecolor{light-yellow}{rgb}{1,0.9,0.8}

\begin{document}

\title{Gravitating superconducting strings with timelike or spacelike currents}

\author{Betti Hartmann $^{(a)}$ }
\email{b.hartmann@jacobs-university.de}

\author{Florent Michel $^{(a),(b)}$}
\email{michel@clipper.ens.fr}

\affiliation{
$(a)$ School of Engineering and Science, Jacobs University Bremen, 28759 Bremen, Germany\\
$(b)$ ICFP, D\'epartement de Physique de l'ENS , 24 rue Lhomond, 75005 Paris, France}
\date\today

\begin{abstract}
We construct gravitating superconducting string solutions of the $U(1)_{\rm local}\times U(1)_{\rm global}$ model
solving the coupled system of Einstein and matter field equations numerically.
We study the properties of these solutions in dependence on the ratio between the symmetry breaking
scale and the Planck mass. Using the macroscopic stability conditions
formulated by Carter, we observe that the coupling to gravity allows for a new stable
region that is not present in the flat space-time limit.
We match the asymptotic metric to the Kasner metric and show that
the relations between the Kasner coefficients and the energy per unit length 
and tension suggested previously are well fulfilled for symmetry breaking scale
much smaller than the Planck mass. 
We also study the solutions
to the geodesic equation in this space-time. While geodesics in
the exterior space-time of standard cosmic strings are just straight lines,
test particles experience a force in a general Kasner space-time and as such bound 
orbits are possible. 

\end{abstract}
\maketitle

\section{Introduction}

Particle physics theories beyond the Standard Model generally predict phase transitions in 
the early universe, during which topological defects can appear \cite{kibble}. 
Since they are, by definition, topologically stable, 
they may survive up to now and have detectable effects. In particular, 
one-dimensional defects, called \textit{cosmic strings} \cite{cosmic_strings,peter_uzan}, appear quite generally and are thought 
to reach a \textit{scaling regime}, 
so that their contribution to the energy density of the universe 
remains finite and constant. 
Due to the fact that these
objects can be extremely heavy they were believed to be 
a possible source of the density 
perturbations that led to structure formation and the anisotropies in the 
cosmic microwave background (CMB) \cite{peter_uzan}. However, the detailed measurements of the 
CMB power spectrum
as obtained by COBE, BOOMERanG and WMAP demonstrated
that cosmic strings cannot be the main source for these anisotropies.
However, in recent years it has been suggested that cosmic strings should generically form at the
end of inflation in inflationary models resulting from String Theory \cite{polchinski}
such as brane inflation \cite{braneinflation}. Moreover, cosmic strings seem to be a generic prediction of supersymmetric 
hybrid inflation \cite{lyth} and grand unified based inflationary models \cite{jeannerot}. 
Even though the origin of these cosmic superstrings is String theory, 
their properties can be investigated in the framework of field theoretical models
\cite{saffin,rajantie,salmi,urrestilla}. 

In general cosmic strings can not end. This allows for two kinds of strings: 
loops and infinitely extended strings. Infinite 
strings are thought to be relatively straight on macroscopic scales and their width is typically 
much smaller than their extent. A string can thus be described in good approximation by 
a one-dimensional object, characterized by quantities integrated over a plane orthogonal to it. 
This procedure is generally not well-defined in curved space-time, but can be used here 
since the metric generated by the string is ``nearly Minkowskian'' reasonably far away from it. 
For infinite straight strings, there are in fact only two relevant quantities: 
the energy per unit length $U$ and the tension $T$. 
In the simplest field theoretical model, namely the Abelian-Higgs model, these quantities are equal \cite{no}.
This can be related to the necessary Lorentz invariance along the string axis. But if neutral currents or some 
microscopic structures (for instance wiggles) are taken into account \cite{witten}, then 
the energy per unit length can be larger than the tension. If the tension
would be larger than the energy per unit length the string would be unstable under transverse perturbations.
The relation between the energy per unit length and the tension of a superconducting string solution
of the $U(1)_{\rm local}\times U(1)_{\rm global}$ model has been discussed in detail in \cite{patrick3} using the formalism 
developed by Carter \cite{carter,carter2}
and it has been suggested that the equation of state is of logarithmic form. 
This has been confirmed numerically in \cite{hartmann_carter}. 

It is also of interest for a possible observation of cosmic strings how the space--time of 
such an object looks like and how test particles would move in this space--time. 
In the case of infinitely thin standard cosmic strings, i.e. cosmic strings without additional
structure that fulfill $U=T$, the space--time is locally flat and geodesics
are just straight lines. However, the space--time is globally conical with deficit angle
$\Delta=8\pi GU$ and as such  light gets deflected by a cosmic string. Next to the observation
of cosmic string signals in the power and polarization spectra of the CMB \cite{cmb_strings1,cmb_strings2,cmb_strings3}
this lensing property of cosmic strings has been suggested to be the prime signature of these objects.
Taking the finite width of
the cosmic string into account \cite{hartmann_sirimachan} bound orbits of massive
test particles are possible, while massless particles can only move on escape orbits. 

The exterior metric of a superconducting string carrying timelike and spacelike currents, respectively, 
has been first discussed in \cite{patrick_denis},
while lensing properties of a 
cosmic string with a lightlike current have been investigated in \cite{patrick2}. The space-time of cosmic strings
with a non-degenerate energy-momentum tensor has been given in \cite{patrick1} and it has been shown that
the parameters in the general Kasner space-time \cite{kramer} can be given in terms of $U$ and $T$. 
This will describe e.g. the exterior of superconducting strings as well as strings with wiggles \cite{Ozdemir:2000qw}.

When considering concrete field theoretical models of gravitating cosmic strings to describe their microscopic properties,
we need to solve the set of coupled Einstein and matter field equations numerically.
This has been done for Abelian-Higgs strings without currents in \cite{clv,brihaye_lubo}. In this paper,
we are interested in solving the full set of coupled Einstein and matter fields equations of a 
$U(1)_{\rm local}\times U(1)_{\rm global}$ model
describing superconducting string solutions with either timelike or spacelike currents in curved space-time.
We will hence be able to determine the metric functions on the full interval from the string axis
out to infinity.  
 
Our paper is organised as follows: in Section II, we discuss the field theoretical
model describing gravitating $U(1)_{\rm local}\times U(1)_{\rm global}$ superconducting strings. 
In Section III, we discuss our numerical results and in particular match
our solutions to the Kasner solutions. In Section IV, we comment on test particle
motion in these space-times and we conclude in Section V.

\section{The Model}
In the following, we will consider the $U(1)_{\rm local}\times U(1)_{\rm global}$ model
in curved space-time. The action reads
\begin{equation}
\label{action}
S=\int d^4 x \sqrt{-g} \left( \frac{1}{16\pi G} R + {\cal L}_{m} \right)
\end{equation}
where $R$ is the Ricci scalar and $G$ denotes Newton's constant. The matter Lagrangian
${\cal L}_{m}$ reads:
\begin{equation}
{\cal L}_{m}=\frac{1}{2}D_{\mu} \phi (D^{\mu} \phi)^*-\frac{1}{4} F_{\mu\nu} F^{\mu\nu}
+\frac{1}{2}\partial_{\mu} \xi (\partial^{\mu} \xi)^*
-V(\phi,\xi)
\end{equation} 
with the covariant derivative $D_\mu\phi=\nabla_{\mu}\phi-ie A_{\mu}\phi$
and the
field strength tensor $F_{\mu\nu}=\nabla_\mu A_\nu-\nabla_\nu A_\mu=\partial_\mu A_\nu-\partial_\nu A_\mu$, 
of the U(1) gauge potential $A_{\mu}$
 with coupling constant $e$. The fields 
$\phi$ and $\xi$ are complex scalar fields with potential
\begin{equation}
V(\phi,\xi)=\frac{\lambda_1}{4}\left(\phi\phi^*-\eta^2_1\right)^2
+\frac{\lambda_2}{4}\xi\xi^*\left(\xi\xi^*-2\eta^2_2\right)
+\frac{\lambda_3}{2} \phi\phi^*\xi\xi^*   \ .
\end{equation}

In cylindrical coordinates $\{t,r,\theta, z\}$ we choose the following Ansatz for the matter fields
\begin{equation}
 A_{\mu} dx^{\mu} = \frac{1}{e}\left(n-P(r)\right)d\theta \ \ , \ \ 
\phi(r,\theta)=\eta_1 h(r) \exp(in\theta) \ \ , \ \ \xi(r,t,z)=\eta_1 f(r) \exp(ikz-i\omega t)
\end{equation}
and
\begin{equation}
\label{metric}
 ds^2= N^2(r) dt^2 - dr^2 - L^2(r) d\theta^2 - K^2(r) dz^2  \ 
\end{equation}
for the metric. 
In the flat space-time limit, i.e. for $G=0$ 
we can define a Lorentz-invariant quantity
${\rm w}:=k^2-\omega^2$ which is typically used to categorize the solutions into the timelike, spacelike
and chiral type for ${\rm w} < 0$, ${\rm w} > 0$ and ${\rm w}=0$, respectively. In the following, 
we will adopt
the viewpoint that we can always go to a suitable frame of reference
to choose either $\omega^2=0$ or $k^2=0$. Hence ${\rm w}=k^2$ and ${\rm w}=-\omega^2$
corresponds to the spacelike and the timelike case, respectively.

\subsection{Equations of motion and boundary conditions}
The dynamics of the metric is given by the Einstein equations which read
\begin{equation}
 R_{\mu\nu}= - 8\pi G \left(T_{\mu\nu} - \frac{1}{2} g_{\mu\nu} T\right) \ ,
\end{equation}
where $T=T^{\sigma}_{\sigma}$ is the trace of the energy-momentum tensor given by
\begin{equation}
 T^{\mu}_{\nu}=\delta^{\mu}_{\nu}{\cal L_{\rm m}} - 2 g^{\mu\sigma} 
\frac{\partial {\cal L_{\rm m}}}{\partial g^{\sigma\nu}}  \ .
\end{equation}

The components of the Ricci tensor then are  \cite{clv}
\begin{equation}
 R^t_t=-\frac{(LKN')'}{LKN} \ \ , \ \ R_r^r = -\frac{N''}{N}-\frac{L''}{L}-\frac{K''}{K} \ \ ,  \ \ 
R_{\theta}^{\theta} = -\frac{(NKL')'}{NKL} \ \ , \ \ R_z^z=-\frac{(NLK')'}{NLK} \ , 
\end{equation}
where here and in the following the prime denotes the derivative with respect to $r$.

We use the rescalings
\begin{equation}
 r\rightarrow \frac{r}{\sqrt{\lambda_1}\eta_1} \ \ , \ \ L\rightarrow \frac{L}{\sqrt{\lambda_1} \eta_1} \ \ , \ \ 
(\omega,k) \rightarrow \sqrt{\lambda_1 }\eta_1 (\omega,k) \ , 
\end{equation}
where $M_{\rm H}=\sqrt{\lambda_1} \eta_1$ corresponds to the mass of the Higgs field. The components of 
the energy-momentum tensor and the field equations will then depend only on the following dimensionless coupling
constants
\begin{equation}
 \alpha:=\frac{e^2}{\lambda_1} \ \ ,  \ \  
\kappa:=8\pi G \eta_1^2 \ \ \ , \ \ \gamma_i:=\frac{\lambda_i}{\lambda_1} \ , \ i=2,3  \ \ .
\end{equation}
The constant $\kappa$ corresponds to the squared ratio between the symmetry breaking scale and
the Planck mass $M_{\rm pl}=G^{-1/2}$. In general, we would expect this to be very small (e.g.
on the order of $10^{-6}$ for GUT scale strings), however, we will typically also
investigate the solutions for higher values of $\kappa$ to understand the general pattern
of solutions.

The components of the energy-momentum tensor (in units of $\lambda_1 \eta_1^4$) are
\begin{eqnarray}
\label{em_super}
 T^t_t&=& \sum\limits_{i=1}^5 \varepsilon_i + u\ \ , \ \ T_r^r=-\varepsilon_1 + \varepsilon_2 - \varepsilon_3 -
\varepsilon_4 + \varepsilon_5 + u \ , \nonumber \\
T_{\theta}^{\theta}&=&\varepsilon_1 - \varepsilon_2 - \varepsilon_3 -
\varepsilon_4 + \varepsilon_5 + u \ \ , \ \ T_{z}^{z}=\varepsilon_1 + \varepsilon_2 + \varepsilon_3 -
\varepsilon_4 - \varepsilon_5 + u \ ,
\end{eqnarray}
where
\begin{equation}
 \varepsilon_1 = \frac{h'^2}{2} + \frac{f'^2}{2} \ \ , \ \ \varepsilon_2 = \frac{h^2 P^2}{2 L^2} \ \ ,  \ \ 
\varepsilon_3= \frac{1}{2\alpha} \frac{P'^2}{L^2} \ ,
\end{equation}
\begin{equation}
 \varepsilon_4=\frac{\omega^2 f^2}{2 N^2} \ \ , \ \ \varepsilon_5=\frac{k^2 f^2}{2 K^2} \ \ , \ \ 
u=\frac{1}{4} (h^2-1)^2 + \frac{\gamma_2}{4} f^2 (f^2 - 2 q^2) + 
\frac{\gamma_3}{2} h^2 f^2  \ .
\end{equation}
The three independent Einstein equations then read
\begin{eqnarray}
 R_t^t &=& -\kappa \left(2\varepsilon_4 + \varepsilon_3 - u \right) \label{eeq1} \ , \\ 
R_{\theta}^{\theta} &=& \kappa \left(2\varepsilon_2 + \varepsilon_3 + u \right) \label{eeq2} \ , \\
R_z^z &=& \kappa \left(2\varepsilon_5 - \varepsilon_3 + u \right) \label{eeq3}
\end{eqnarray}
subject to the constraint (which is not independent)
\begin{equation}
\label{constraint}
 R^r_r=\kappa\left(2\varepsilon_1 + \varepsilon_3 + u \right)  \ .
\end{equation}
The Euler-Lagrange equations which result from the variation of the action
with respect to the matter fields are
\begin{equation}
 \label{el1}
\frac{(LKNh')'}{LKN} = h(h^2-1) + \gamma_3 h f^2 + \frac{h P^2}{L^2} \ , 
\end{equation}
\begin{equation}
 \label{el2}
\frac{(LKNf')'}{LKN} = \frac{k^2 f}{K^2} - \frac{\omega^2 f}{N^2} + \gamma_3 h^2 f + \gamma_2 f (f^2 -q^2) \ , 
\end{equation}
\begin{equation}
 \label{el3}
\frac{L}{NK}\left(\frac{NK P'}{L}\right)' = \alpha P h^2 \ .  
\end{equation}
Equations (\ref{eeq1})-(\ref{eeq3}) and (\ref{el1})-(\ref{el3}) have to be solved subject to 
appropriate boundary conditions. At $r=0$ the requirement of regularity and the fact that we would like string-like
solutions leads to
\begin{equation}
 \label{bc1}
h(0)=0 \ \ , \ \ f'(0)=0 \ \ , \ \ P(0)=n \ \ , \ \ L(0)=0 \ \ ,  \ \ L'(0)=1 \ \ , \ \ N(0)=1 \ \ , \ \ 
N'(0)=0 \ \ , \ \ K(0)=1 \ \ , \ \ K'(0)=0 \ .
\end{equation}
Finiteness of energy requires that
\begin{equation}
\label{bc2}
 h(\infty)=1 \ \ , \ \ f(\infty)=0 \ \ , \ \ P(\infty)= 0  \ .
\end{equation}

In our numerical calculations, we have solved the coupled system (\ref{eeq1})-(\ref{eeq3}), (\ref{el1})-(\ref{el3})
subject to the boundary conditions (\ref{bc1}), (\ref{bc2}). In all our calculations, we have
checked that (\ref{constraint}) is fulfilled.

\subsection{Energy per unit length, tension and current}
In \cite{patrick1} the energy per unit length $U$ and tension $T$ have been used as macroscopic parameters
to describe superconducting string solutions in the $U(1)_{\rm local}\times U(1)_{\rm global}$ model in flat space-time
and to investigate the stability of these objects.
The corresponding expressions in curved space-time are a straightforward generalization 
of that and read
\begin{equation}
 U=\int\int \sqrt{-h}  T^t_t \ dr d\theta= 2\pi \int \int L \ T^t_t \ dr \ \ , \ \ 
T=-\int \sqrt{-h} \ T^z_z \ dr  d\theta = -2\pi \int L \ T^z_z \ dr ,
\end{equation}
where $h$ corresponds to the determinant of the induced metric on the $(t,z)$-plane. 
It is also possible to define the so-called Tolman mass of these solutions which corresponds
to the gravitational active mass (see e.g. \cite{brihaye_lubo}), but we would like
to compare our results to the flat space-time limit studied in \cite{patrick3} and
hence define $T$ and $U$ as given above. Furthermore, the Noether current
associated to the unbroken U(1) symmetry reads
\begin{equation}
J^{\mu} =  \frac{i}{2} \left (\xi^* \nabla^{\mu} \xi - \xi \nabla^{\mu} \xi^*\right) \ ,
\end{equation}
which has non-vanishing, dimensionless components
\begin{equation}
 J^{t} = \omega \frac{f^2}{N^2} \ \ , \ \ J^{z}=k\frac{f^2}{K^2}  \ .
\end{equation}
This current is covariantly conserved $\nabla_{\mu} J^{\mu}=0$, which
implies $\partial_{\mu} \left(\sqrt{-g} J^{\mu}\right)=0$ and since
$\sqrt{-g}$ is independent of $(t,z)$ we have $\partial_{\mu} J^{\mu}=0$. 
Finally, we can also define the charge number density (in analogy to \cite{patrick3})
which reads
\begin{equation}
 J=2\pi  \int  \sqrt{\left\vert J^t J_t + J^z J_z\right\vert} \ L \ dr  \ .
\end{equation}
In \cite{carter} a macroscopic stability condition for superconducting strings
has been suggested. This requires that the propagation speeds of transverse (T) and longitudinal (L)
perturbations, respectively, should both be real for the string to be stable.
The squared propagation speeds are given by
\begin{equation}
 c_{\rm T}^2 = \frac{T}{U} \ \ \ , \ \ \ c_{\rm L}^2 = -\frac{dT}{dU} \ 
\end{equation}
and should hence both be positive. 
Since our definitions of $T$ and $U$ are covariantly defined expressions and the original 
work of Carter was formulated in a fully covariant way \cite{carter,carter2}, we 
expect these criteria to also hold in the gravitating case and use them in the following
to decide about the stability of gravitating superconducting strings.

\subsection{Asymptotic behaviour of the metric functions}
Far away from the string core where the matter functions have reached their vacuum
values, we would expect that the metric functions have the behaviour of those
of a general cylindrically symmetric vacuum space-time given by the Kasner metric
which has the following form 
\begin{equation}
\label{metric_general}
 ds^2 =  \left( \frac{r}{r_\sigma} \right)^{2a} dt^2 - dr^2 - \left( \frac{r}{r_\sigma} \right)^{2 b} dz^2 
- \gamma r^2 \left( \frac{r}{r_\sigma} \right)^{2 c} d\theta^2  \ ,
\end{equation}
 where $a$, $b$ and $c$  are real coefficients subject to the Kasner conditions:
 \begin{equation}
 a + b + c = 0 \ \ , \  \   a^2 + b^2 + (c+1)^2 = 1  \ ,
 \end{equation}
while $r_\sigma$ can be thought of as the typical radius of the string. $\gamma$ is an additional
parameter that  determines the deficit angle of the space-time $\Delta$.
Note that the first Kasner condition can be used to eliminate $c$ in the second one.  
We then get a second-order polynomial in $b$, which has real solutions if and only if 
$a \in \left[-\frac{1}{3},1 \right]$. Then $b$ is given by:
\begin{equation}
\label{abc}
b = \frac{1-a \pm \sqrt{-3 a^2 + 2 a +1}}{2} \equiv b^\pm \ \ , \ \ 
c =- \frac{1+a \pm \sqrt{-3 a^2 + 2 a +1}}{2} \equiv c^\pm
\end{equation}
so that $c^\pm =  b^\mp-1$. 

In \cite{patrick1} it was argued that with the identification
\begin{equation}
 a=-b=2G (U-T) + {\cal O}(G^2) \ \ , \ \ c={\cal O}(G^2) \ \ , \ \ \gamma=1-4G(U+T)\
\label{parameters}
\end{equation}
the Kasner metric would describe the outside of
a general cosmic string with energy per unit length $U$ and tension $T$ under the assumption
that the string is infinite, straight and has negligible width. The deficit angle of the space-time
is then given by $\Delta = 4\pi G (U+T)$. 

For ``standard'' cosmic strings we have $U=T$ and the metric describes a conical
space-time with deficit angle $\Delta= 8\pi G U$ \cite{cosmic_strings}.
In this case, the space-time is locally flat and geodesics are just straight lines.
Since the space-time has a deficit angle, gravitational lensing appears, but planetary, i.e.
bound orbits are not possible. Note that this changes
when taking the finite core of the cosmic string into account \cite{hartmann_sirimachan}.

\section{Numerical results}
The solutions to the equations (\ref{eeq1})-(\ref{eeq3}) and (\ref{el1})-(\ref{el3}) 
are only known numerically. We have solved these equations
using the ODE solver COLSYS \cite{colsys}. 
The solutions have relative errors on the order of $10^{-6}-10^{-10}$.
In the following, we have restricted our analysis to the case $n=1$ unless otherwise
stated. 

\subsection{General behaviour and Kasner coefficients}

\begin{figure}[t]
\begin{center}
\subfigure[][matter functions $P(r)$, $h(r)$, $f(r)$]
{\label{matter_electric}\includegraphics[width=8.0cm]{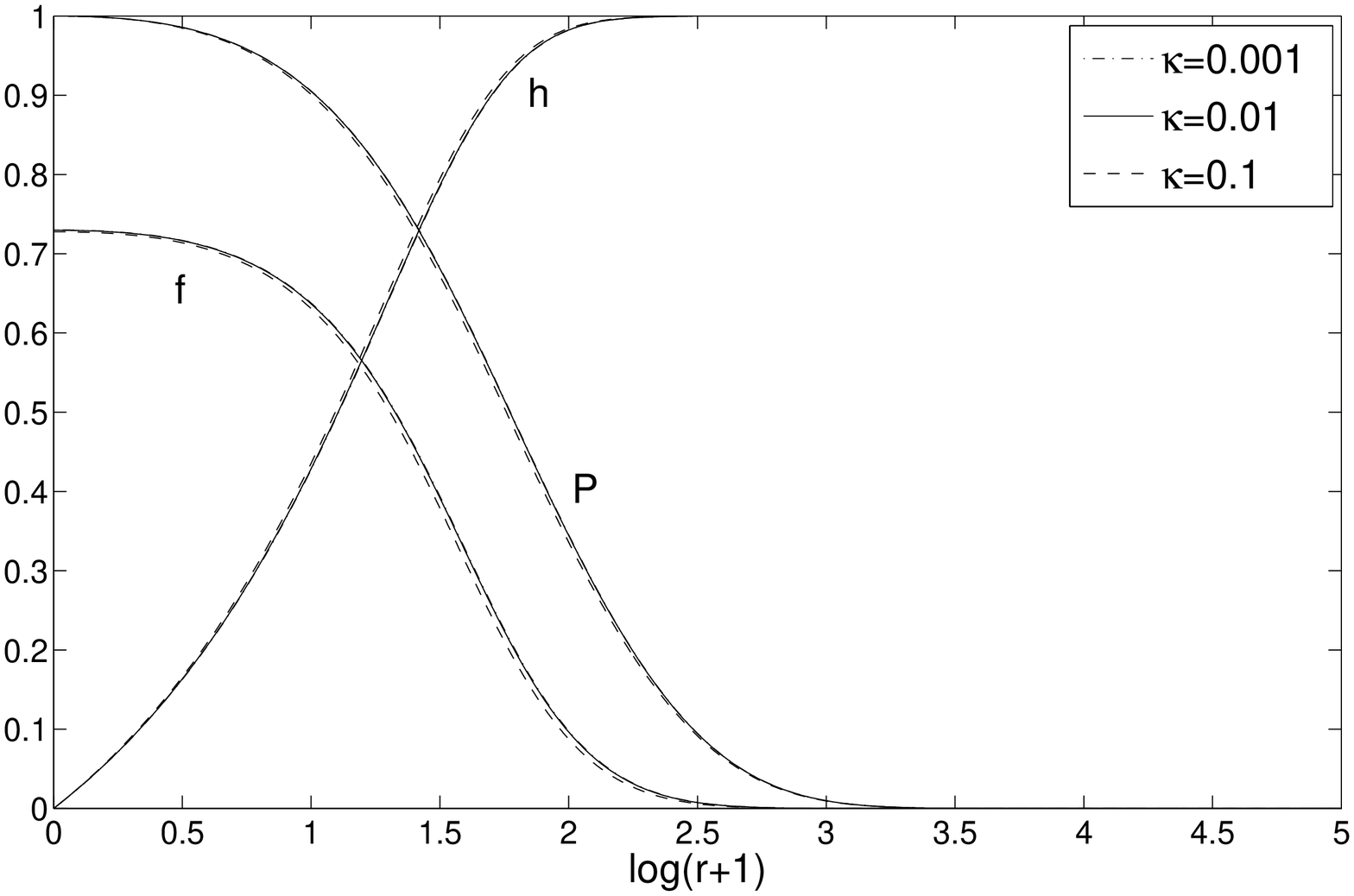}}
\subfigure[][metric function $N(r)$]{\label{electric_N}\includegraphics
[width=8.0cm]{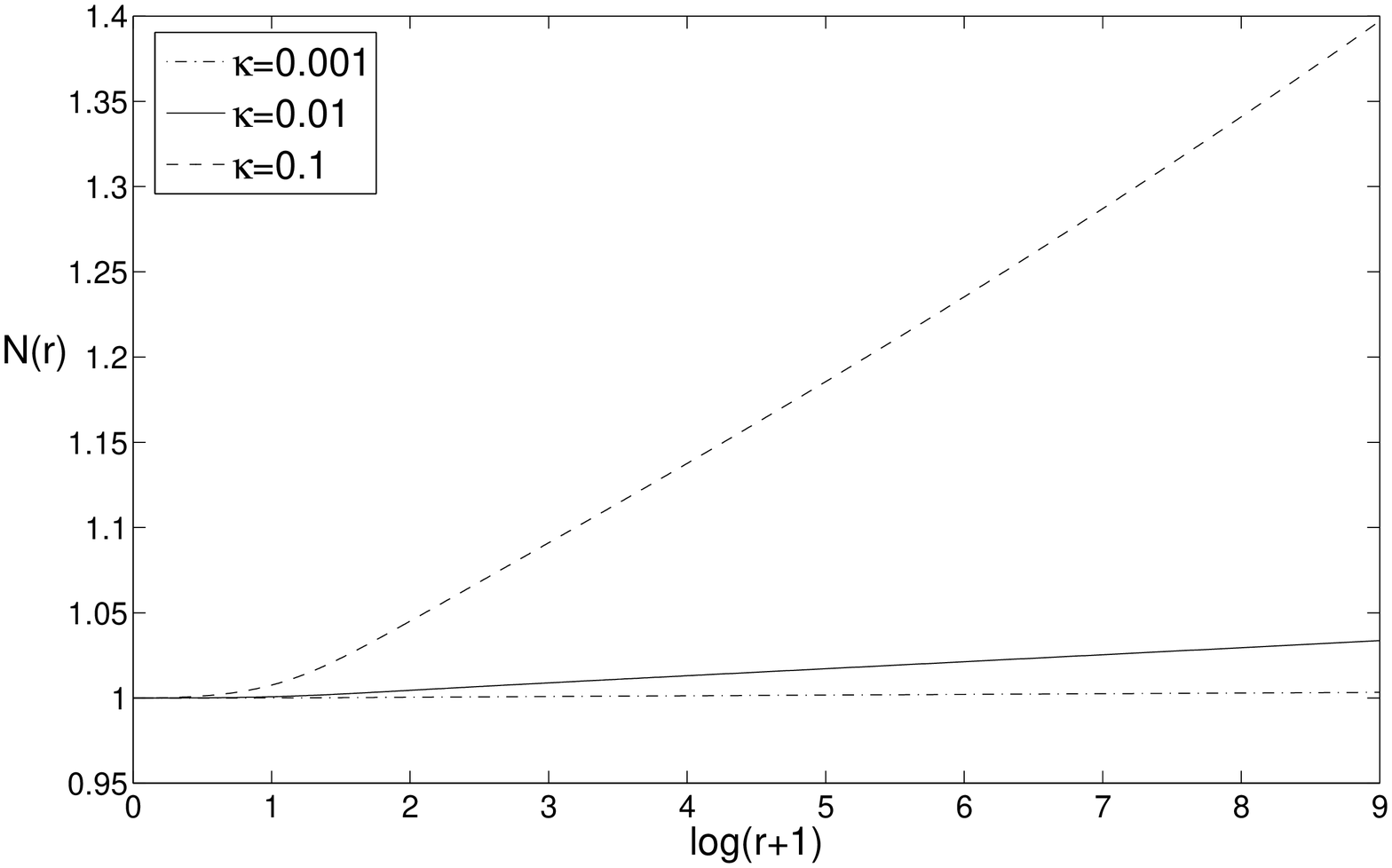}}\\
\subfigure[][metric function $L(r)$]{\label{electric_L}\includegraphics[width=8.0cm]
{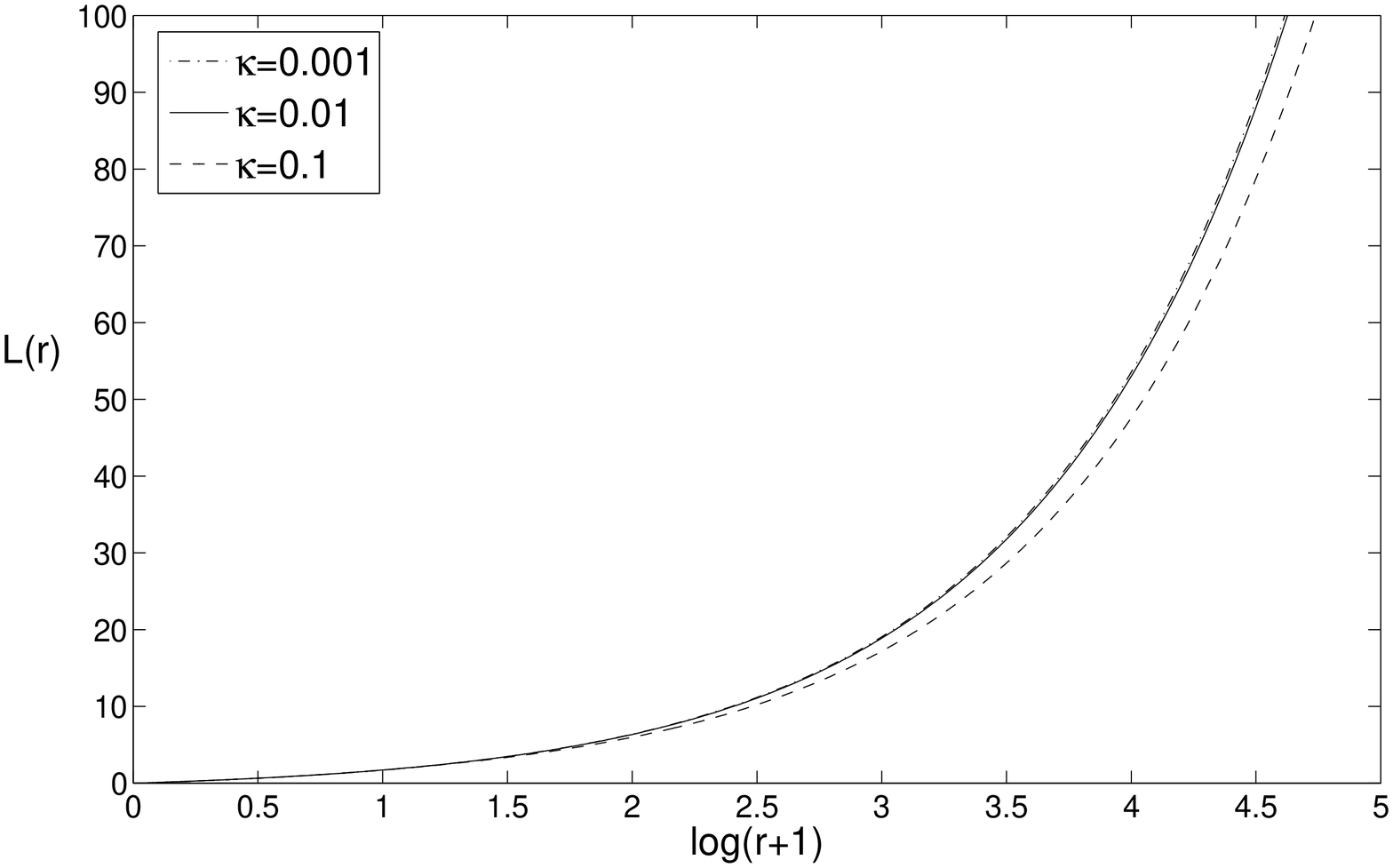}}
\subfigure[][metric function $K(r)$]{\label{electric_K}\includegraphics
[width=8.0cm]{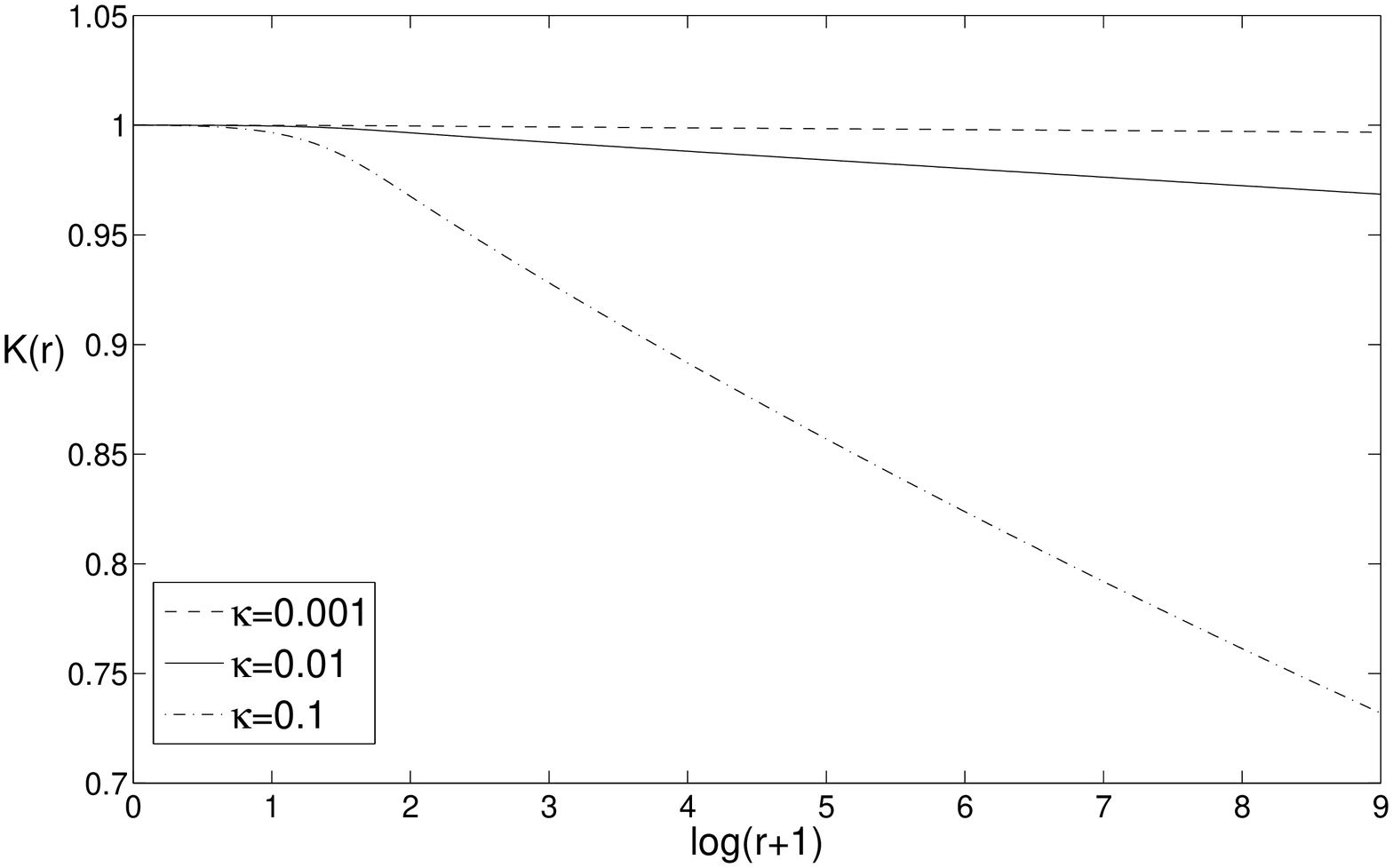}}
\end{center}
\caption{\label{electric} The matter and metric functions for a gravitating
superconducting string with ${\rm w}=-0.3$, 
$\gamma_2=3$, $\gamma_3=2$, $q=0.7$, $\alpha=0.1$ and different values of $\kappa$.}
\end{figure}

In Fig.\ref{electric} we show the matter and metric functions of a solution for
${\rm w}=-0.3$, 
$\gamma_2=3$, $\gamma_3=2$, $q=0.7$, $\alpha=0.1$ and three different values of $\kappa$.
As is apparent, the matter functions vary only very little (the different
cases are barely distinguishable on the plot), however,
the metric functions change quite strongly. 
When increasing $\kappa$ the metric functions $N(r)$ and $K(r)$ deviate more strongly from their
flat space-time values $N(r)=K(r)\equiv 1$, while the slope of $L(r)$ at large $r$ decreases
with increasing $\kappa$ signaling --as expected -- an increase in the deficit angle of the space--time.

The outside space-time of a superconducting string
should be given by the Kasner space-time with a specific relation between the Kasner coefficients
and $U$ and $T$ (see (\ref{parameters})) \cite{patrick1}.
We observe that our numerical solutions indeed
possess an asymptotic behaviour 
governed by the Kasner metric (\ref{metric_general}). Matching the space-time of the solution
given in Fig.\ref{electric} at large $r$ with the Kasner metric we find $a=0.00040092$, 
$b=-0.00040076$, 
$c=-0.00000054$ for $\kappa=0.001$ and $a=0.04098961$, $b=-0.03937305$, 
$c=-0.00165716$ for $\kappa=0.1$, respectively.

\begin{figure}
\centering
\epsfysize=7cm
\mbox{\epsffile{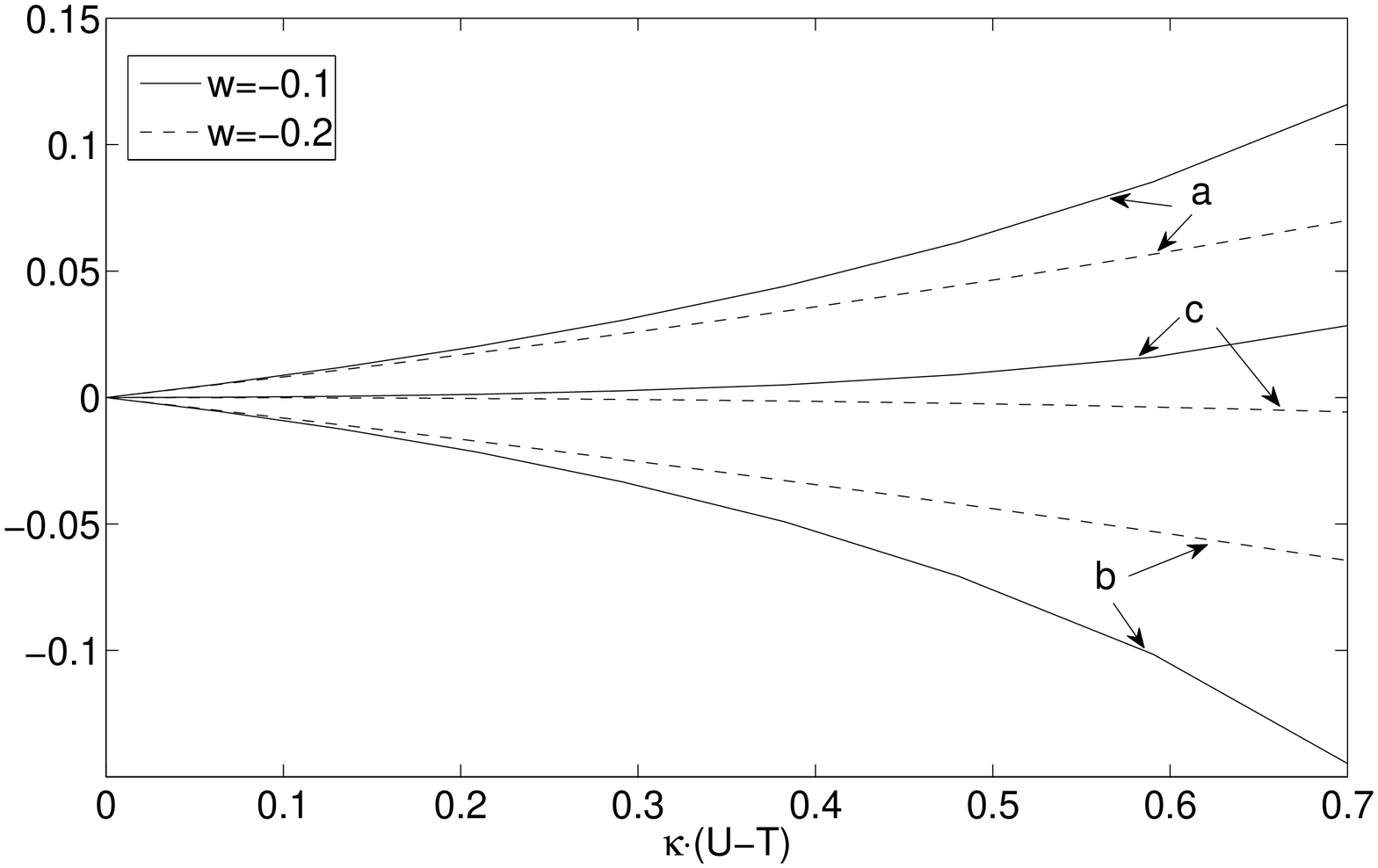}}
\caption{\label{kasner_coeffn}
We show the value of the Kasner coefficients $a$, $b$ and $c$ for gravitating
superconducting string solutions with timelike current and 
$\gamma_2=3$, $\gamma_3=1.74$, $q=0.69$, $\alpha=0.1$ in dependence on $\kappa(U-T)$.
We give the coefficients for two different values of ${\rm w}$.}
\end{figure}

In Fig.\ref{kasner_coeffn} we give the values of the Kasner coefficients in dependence
on $\kappa (U-T)$ for a superconducting string with timelike current (the results are qualitatively 
similar in the spacelike
case, this is why we do not present them here) and
$\gamma_2=3$, $\gamma_3=1.74$, $q=0.69$, $\alpha=0.1$.
For small values of $\kappa$ we recover the behaviour given in
(\ref{parameters}). For larger values of $\kappa$ the parameter $c$ starts to deviate from zero, while
$a$ and $b$ no longer depend linearly on $\kappa(U-T)$. We hence conclude that for small
values of $\kappa(U-T)$ the description of the metric outside the superconducting 
string by a Kasner metric with the identification
(\ref{parameters}) is a very good approximation. However, with our techniques, we are also able to
determine the behaviour of the metric functions inside the string core up to the string
axis at $r=0$. This is only possible
when solving the equations of motion numerically. In the following, we want to discuss
the macroscopic stability of gravitating superconducting strings. For this, we will need
to integrate the matter and metric functions from $r=0$ to $r=\infty$. It is hence
crucial to know the matter and metric functions on the full interval $r\in [0:\infty[$.

\subsection{Macroscopic stability}
\begin{figure}
\centering
\epsfysize=7cm
\mbox{\epsffile{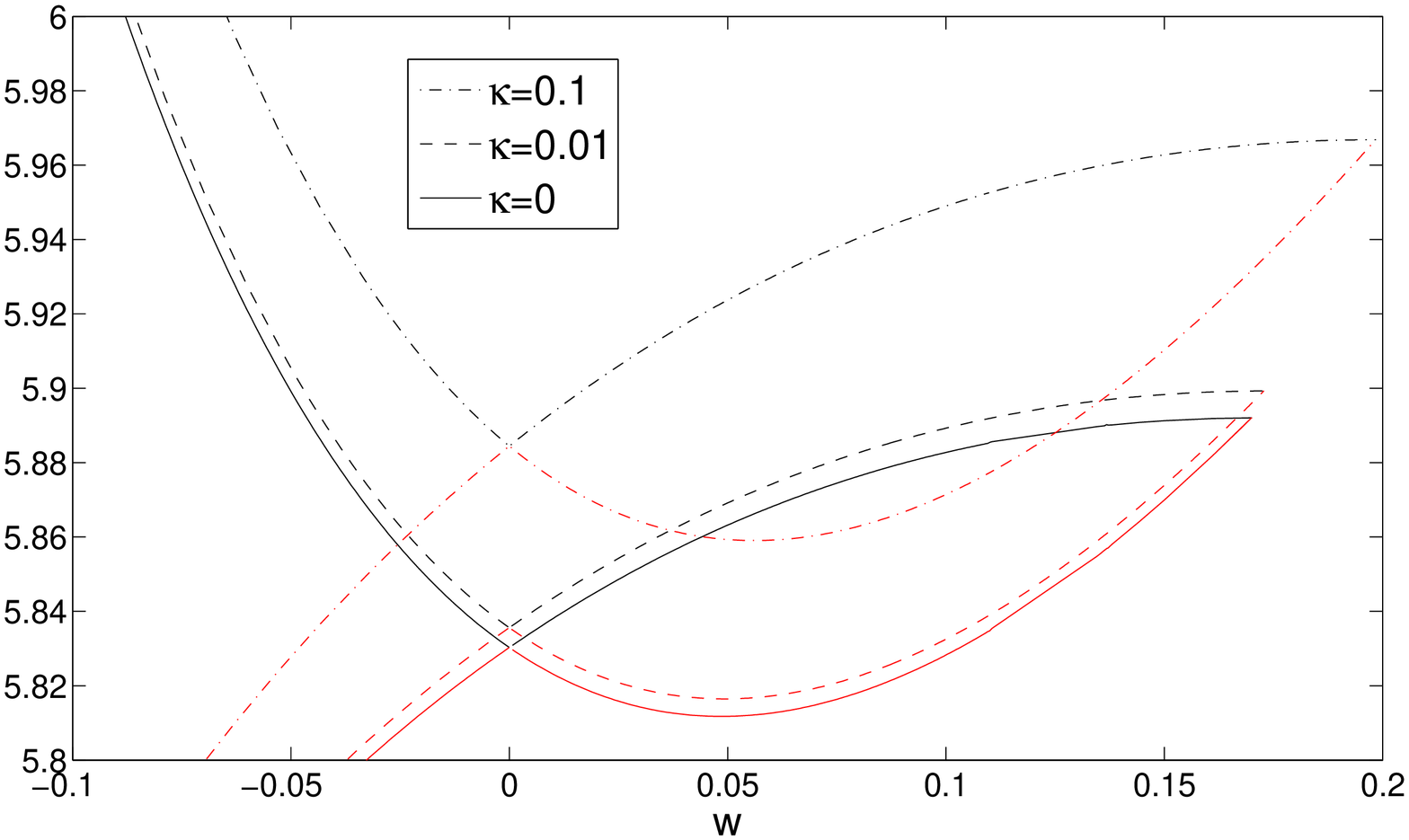}}
\caption{\label{TU_omega}
We show the energy per unit length $U$ (black) as well as the tension $T$ (red) as function
of ${\rm w}=k^2-\omega^2$ for
superconducting strings for different values of $\kappa$ and $\gamma_2=3$, $\gamma_3=2$, $q=0.7$
and $\alpha=0.1$.}
\end{figure}

In flat space-time it was found \cite{patrick3} that in the spacelike regime ${\rm w} > 0$ the energy per unit length
$U$ and the tension $T$ are always positive and hence $c_{\rm T}^2 > 0$. However, while
the energy per unit length is an increasing function of ${\rm w}$ in the spacelike regime,
the tension $T$ decreases only for ${\rm w}$ close to the chiral limit ${\rm w}=0$. For sufficiently
small ${\rm w}$ it was hence found that the strings are stable with $c_{\rm L}^2 > 0$, while
for larger values $c_{\rm L}^2 < 0$.  In the timelike regime, the energy per unit length
and tension diverge at the approach of the phase frequency threshold which corresponds to the
value of $-{\rm w}$ equal to the mass of the scalar boson. 
In the following, we have fixed $\gamma_2=3$, $\gamma_3=2$, $\alpha=0.1$, $q=0.7$ unless otherwise stated.
This choice of parameters fulfills all the requirements such that the local  $U(1)$ symmetry is broken and the global
$U(1)$ symmetry remains unbroken in the flat space-time limit. In the following, we will be interested in the way that the energy
per unit length $U$, the tension $T$ as well as the charge number density $J$ change with $\kappa$. In Fig.\ref{TU_omega}
we plot $U$ and $T$ as function of ${\rm w}$ for ${\rm w}$ close to ${\rm w}=0$. For $\kappa=0$ we recover the results
given in \cite{patrick3}.

\begin{figure}
\centering
\epsfysize=7cm
\mbox{\epsffile{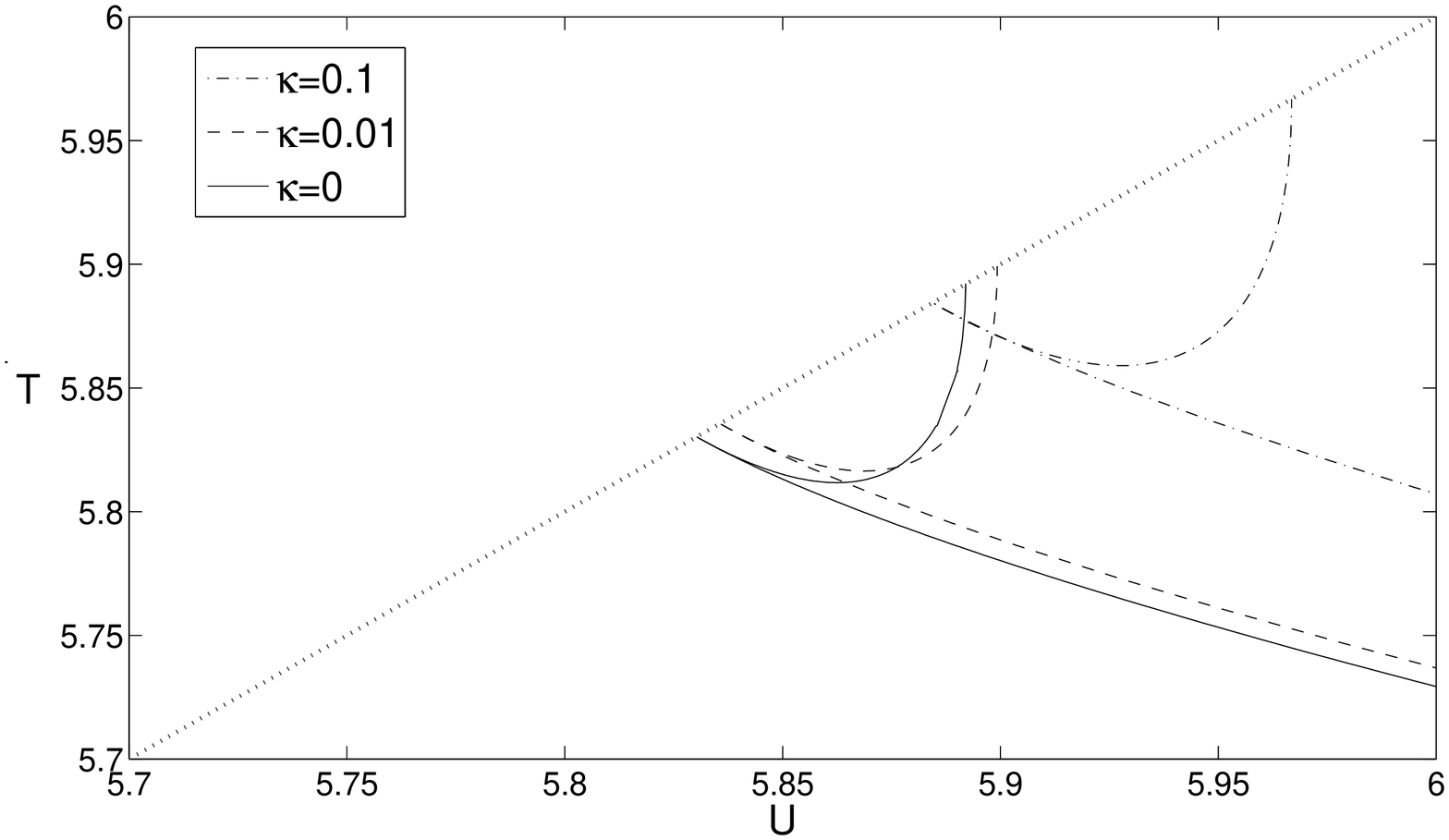}}
\caption{\label{UvsT}
We show the tension $T$ as function of the energy per unit length $U$ for
superconducting strings for different values of $\kappa$ and $\gamma_2=3$, $\gamma_3=2$, $q=0.7$
and $\alpha=0.1$.}
\end{figure}

\begin{figure}
\centering
\epsfysize=7cm
\mbox{\epsffile{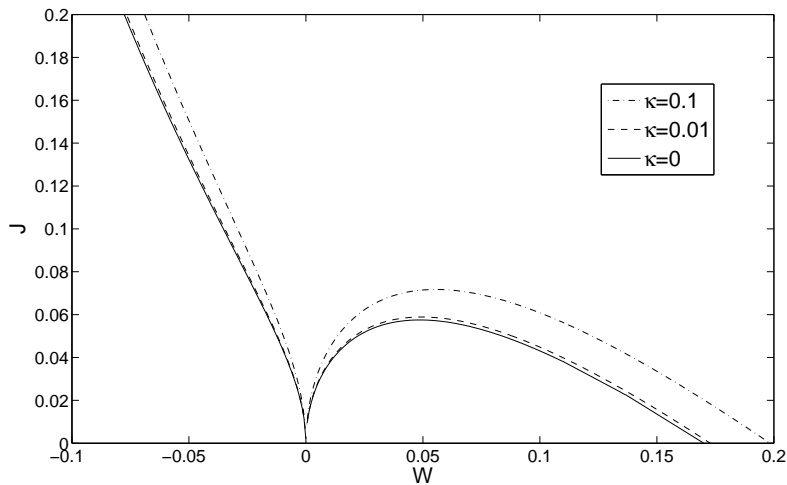}}
\caption{\label{current}
We show the charge number density $J$ as function of ${\rm w}=k^2-\omega^2$ for
superconducting strings for different values of $\kappa$ and $\gamma_2=3$, $\gamma_3=2$, $q=0.7$
and $\alpha=0.1$.}
\end{figure}

Note that we plot the dependence on ${\rm w}$ and not on ${\rm sgn(w)}\sqrt{\rm \vert w\vert}$. We do not
present our results in dependence on this latter quantity here, but have convinced ourselves that the plots
look qualitatively similar to those presented in \cite{patrick3}. We observe that the main features are still present
for $\kappa\neq 0$. The value of ${\rm w}={\rm w}_{\rm cr}^{0}$ at which $T=U$ in the spacelike regime corresponds
to the value of ${\rm w}$ where $f(0)=0$ and hence $f(r)\equiv 0$. We find that ${\rm w}_{\rm cr}^{0}$
increases with $\kappa$, as do the corresponding values of $U$ and $T$. This leads also to the observation
that the range of ${\rm w}$ in the spacelike regime for which $c_{\rm L}^2 > 0$ increases 
since the minimal value of $T$ appears
at larger values of ${\rm w}$. Hence, the coupling to gravity enhances the interval of ${\rm w}\geq 0$ in which strings are stable.
This is also seen in Fig.\ref{UvsT}, where we plot $T$ as function of $U$. Obviously, $dT/dU < 0$ for ${\rm w}$ close to the
chiral limit ${\rm w}=0$. We also give the charge number density $J$ in Fig.\ref{current}. This shows
that this quantity vanishes at ${\rm w}_{\rm cr}^0$ as well as at ${\rm w}=0$. In the spacelike regime the charge number density
first rises reaching a maximum at ${\rm w}_{\rm max}$ and then decreases again to zero at ${\rm w}_{\rm cr}^0$. 
We find that ${\rm w}_{\rm max}$ increases with $\kappa$. In the timelike regime the charge number density
rises strongly and the bigger $\kappa$ the bigger $J$ for a given value of ${\rm w}$.

\begin{figure}
\centering
\epsfysize=7cm
\mbox{\epsffile{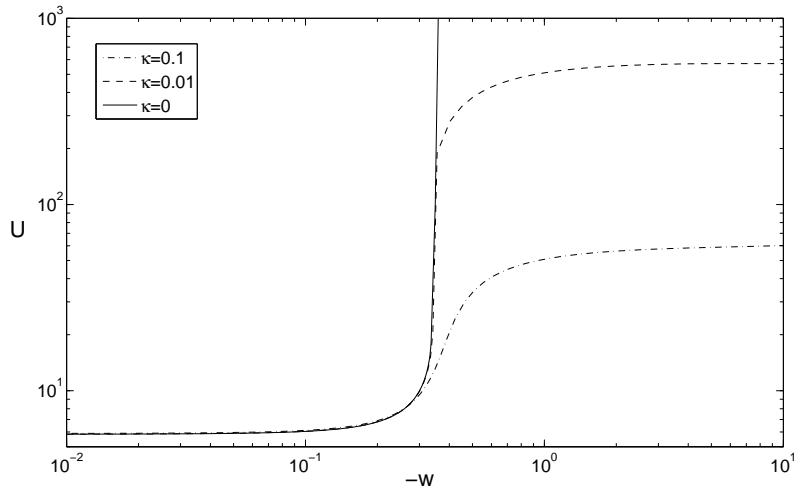}}
\caption{\label{u_om_electric}
We show $U$  as function of $-{\rm w}$ for different values of $\kappa$ and $\gamma_2=3$, $\gamma_3=2$, $q=0.7$
and $\alpha=0.1$.
}
\end{figure}

\begin{figure}
\centering
\epsfysize=7cm
\mbox{\epsffile{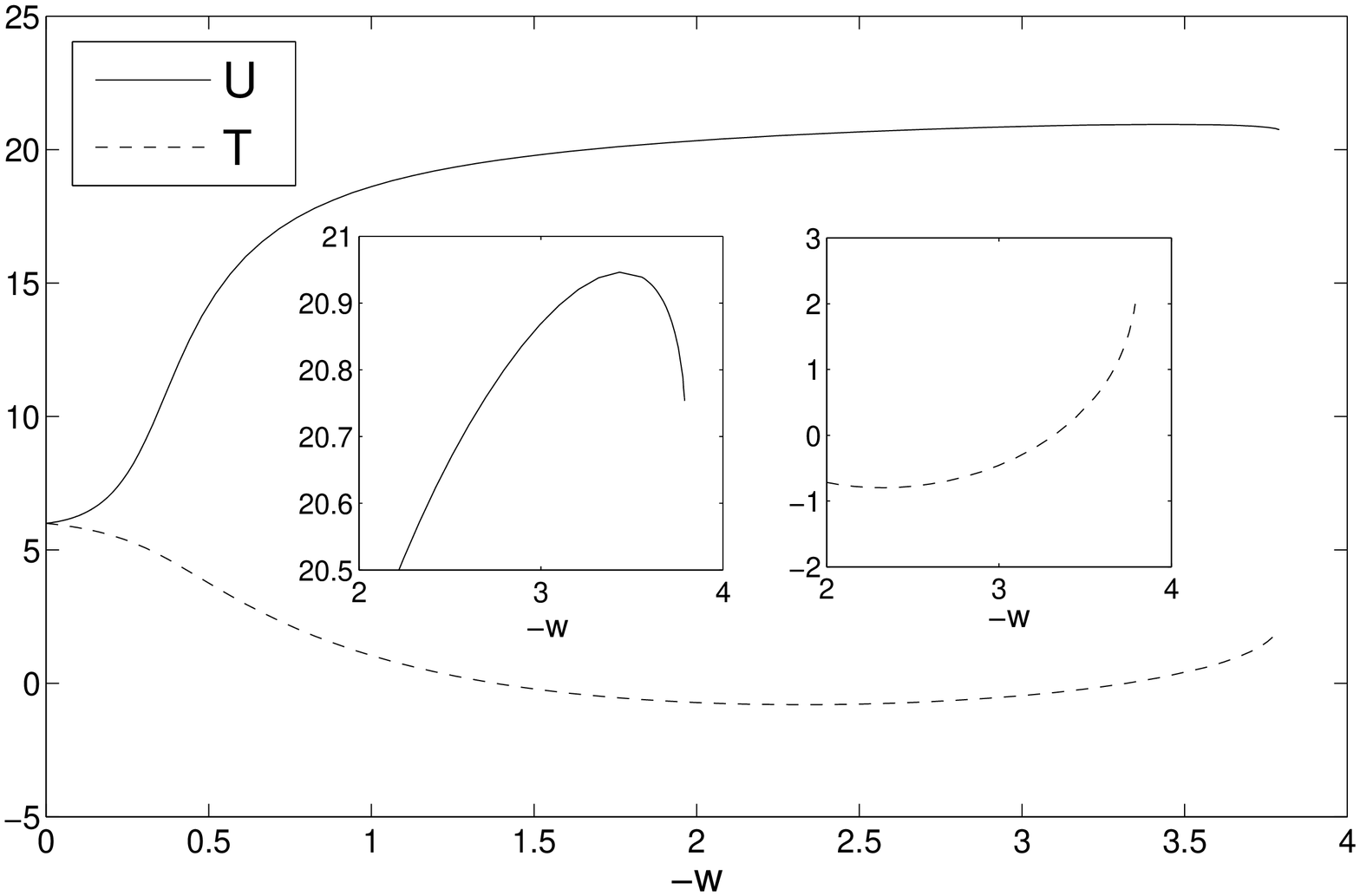}}
\caption{\label{UTelectric_stable}
We show $U$ and $T$ as function of $-{\rm w}$ for $\kappa=0.3$  and $\gamma_2=3$, $\gamma_3=2$, $q=0.7$
and $\alpha=0.1$. The small subfigures show the energy per unit length $U$ and the tension $T$, respectively
at large $-{\rm w}$. Clearly, $U$ decreases, while $T$ increases and becomes 
positive for sufficiently large $-{\rm w}$ signaling the stability of the superconducting string. }
\end{figure}

\begin{figure}
\centering
\epsfysize=7cm
\mbox{\epsffile{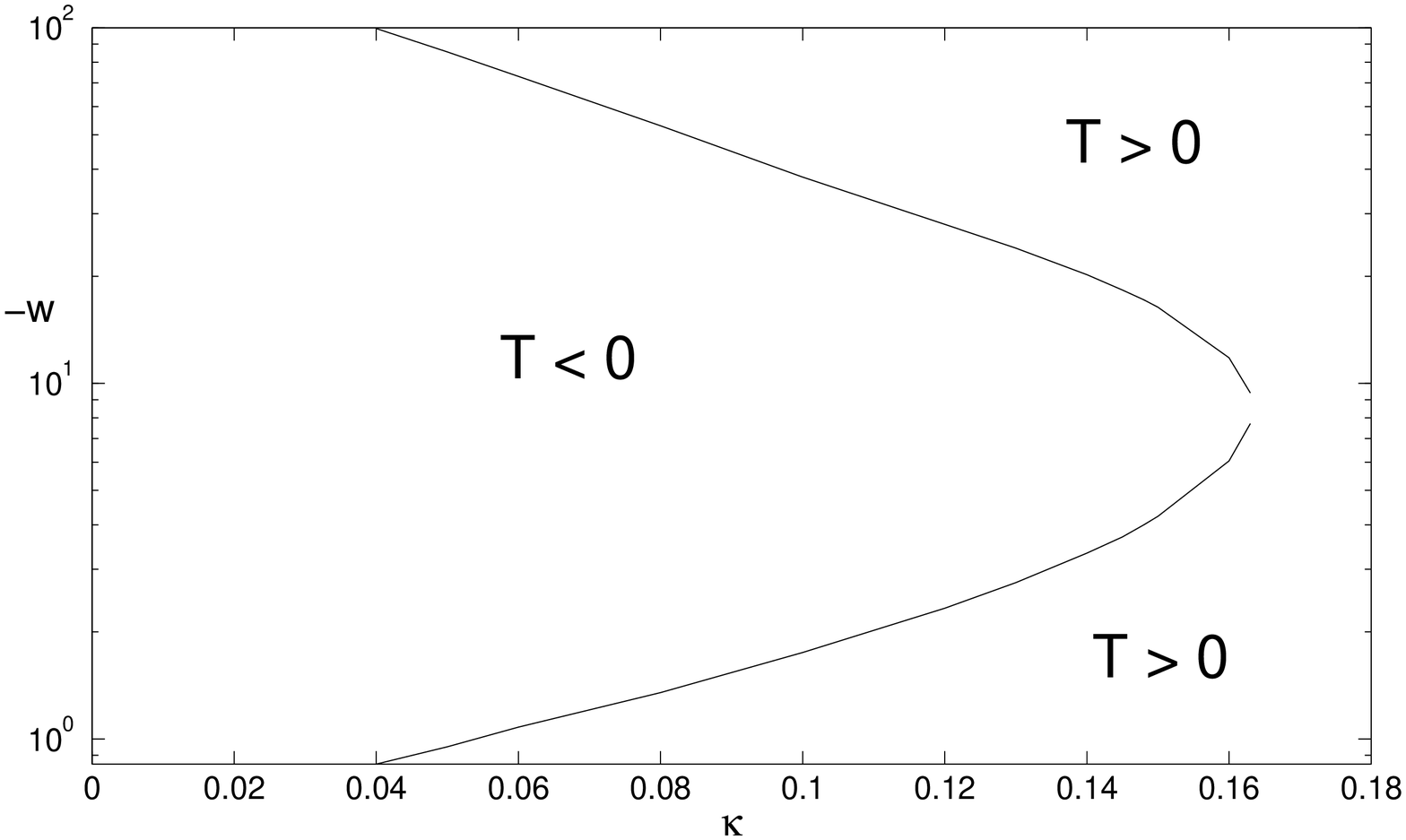}}
\caption{\label{Tnegative}
We show the region in the ${\rm w}$-$\kappa$ domain for which $T < 0$ (and hence $c_T^2 < 0$).
Note that for $\kappa=0$ the tension $T > 0$ in the spacelike regime. This changes when considering
the superconducting string in curved space-time for sufficiently small, but non-vanishing $\kappa$.
This is for $\gamma_2=3$, $\gamma_3=2$, $q=0.7$
and $\alpha=0.1$.}
\end{figure}

In the timelike regime we find in analogy with \cite{patrick3} 
that there exists a phase frequency threshold
at which $U\rightarrow +\infty$ and $T\rightarrow -\infty$ for $\kappa=0$. This corresponds to the 
value of ${\rm w}$ at which scalar bosons
can be produced. Interestingly, we observe that this phase frequency threshold seems to be absent when $\kappa\neq 0$.
This is shown in Fig.\ref{u_om_electric}, where we give $U$ as function of ${\rm w}$ in the timelike regime. Clearly, for large values
of $-{\rm w}$ the energy per unit length tends to a finite value for $\kappa\neq 0$. Interestingly, we find that the coupling
to gravity can hence stabilize the strings. This is shown in Fig.\ref{UTelectric_stable}, where we present $U$ and $T$ in the timelike
regime for $\kappa=0.3$. In an interval close to ${\rm w}=0$ we find that both $U$ and $T$ are positive with $U$ increasing
and $T$ decreasing such that $c_{\rm L}^2 > 0$ and $c_{\rm T}^2 > 0$. Hence, the strings are stable. Decreasing ${\rm w}$ further leads
to $T$ becoming negative such that while $c_{\rm L}^2 > 0$ we now have $c_{\rm T}^2 < 0$ and the strings are unstable. However, in
contrast to the flat space-time case, where $c_{\rm T}^2 < 0$ up to the phase frequency threshold, we observe that for sufficiently
small ${\rm w}$ the tension becomes positive again. As such $c_{\rm T}^2 > 0$. Furthermore, for these
values of ${\rm w}$ we find that $U$ decreases, while $T$ increases such that $c_{\rm L}^2 > 0$ and the
strings are macroscopically stable. 
We conclude that there are 
hence two stable
regions in the timelike regime for $\kappa\neq 0$. Hence, gravity can stabilize the strings for large values of the current.
To get an idea how the parameter ranges in which the string becomes stable depend on $\kappa$, we show the values
of ${\rm w}$ where $T=0$ in dependence on $\kappa$ in Fig.\ref{Tnegative}. In between the two curves the tension $T$ is negative
and hence $c_{\rm L}^2$ is negative. We observe that the value of ${\rm w}$ close to ${\rm w}=0$ at which $T$ vanishes and $dT/d{\rm w} < 0$
increases with increasing $\kappa$. On the other hand, the value of ${\rm w}$ at which $T$ vanishes and
$dT/d{\rm w} > 0$ decreases with increasing $\kappa$. Hence, the range ${\rm w}$ for which $c_{\rm T}^2 < 0$ decreases with
increasing $\kappa$. Furthermore, our numerical results indicate that $c_{\rm T}^2 > 0$ for all ${\rm w}$ if $\kappa$ is sufficiently
large. For $\gamma_2=3$, $\gamma_3=2$, $\alpha=0.1$, $q=0.7$ we find that this happens at $\kappa\approx 0.16$.

\section{Motion of test particles}

In order to probe space-times it is crucial to understand how test particles move in these. We here consider structure-less, point-like
particles that move on geodesics in the gravitational field of a superconducting string. If we consider the numerically given 
space-time in terms of the metric functions $N$, $L$ and $K$ we have to solve the geodesic equation numerically. This approach has
been taken in \cite{hartmann_sirimachan} for standard cosmic strings and it was found that the finite core width of the string
allows for additional bound orbits of massive test particles close to the string core. Since the string core
and hence also the radius of the orbits are on the same order as the inverse scalar boson mass these cannot account for planetary orbits.
On the other hand, the string could be ``dressed'' by massive particles and this could be important for gravitational
wave emission. However, our numerical results indicate that the power of this emission would be rather small. This is why we do not
investigate this further here.

As shown above, we find that the metric outside the string core is well matched by the Kasner space-time. Since this space-time
can be given analytically we are able to make some 
general statements about the geodesic motion.

Since the observation of gravitational lensing by a cosmic string has been suggested as a possibility
to detect these objects directly, we discuss the motion of massless test particles in the Kasner space-time
in the following.
We also comment on the motion of massive test particles.

\subsection{The geodesic equation}

The Kasner metric has three obvious Killing vectors: $\partial_t$, $\partial_\theta$ and $\partial_z$. The three associated constants of motion are:
\begin{equation} \label{eq1}
P_t=-\left( \frac{r}{r_\sigma} \right)^{2 a} \frac{dt}{d \lambda} \ \ , \ \ 
P_z=\left( \frac{r}{r_\sigma} \right)^{2 b} \frac{dz}{d \lambda} \ \ , \ \ 
P_\theta= \gamma r^2 \left( \frac{r}{r_\sigma} \right)^{2 c} \frac{d\theta}{d \lambda}  \ ,
\end{equation}
where $\lambda$ is an affine parameter that can be identified with proper time for time-like geodesics.
$P_t$ corresponds to the energy, $P_z$ to the linear momentum in the $z$-direction and
$P_{\theta}$ to the angular momentum of a test particle moving in this space-time, respectively.
The geodesic Lagrangian then reads
\begin{equation}
 {\cal L}_{\rm geo} = \frac{1}{2} g_{\mu\nu} \frac{dx^{\mu}}{d\lambda} \frac{dx^{\nu}}{d\lambda} = 
\frac{1}{2}\epsilon  \ ,
\end{equation}
where $\epsilon$ takes the value $0$ and $1$ for massless and massive particles, respectively.
Using the constants of motion (\ref{eq1}) the equation of motion for $r$ reads
\begin{equation}
P_t \frac{dt}{d \lambda} + P_z \frac{dz}{d \lambda} 
+ P_\theta \frac{d\theta}{d \lambda} + \left( \frac{dr}{d\lambda} \right)^2 = -\epsilon
\end{equation}
which can be rewritten as follows
\begin{equation}\label{eq4}
\left( \frac{dr}{dt} \right)^2 + \left[ -\left( \frac{r}{r_\sigma} \right)^{2 a} + 
\frac{\epsilon}{P_t^2} \left( \frac {r}{r_\sigma} \right)^{4 a} +\left( \frac{P_z}{P_t} \right)^2  
\left( \frac{r}{r_\sigma} \right)^{4 a - 2 b} + \frac{1}{\gamma r^2} 
\left( \frac{r}{r_\sigma} \right)^{4 a - 2 c} \left( \frac{P_\theta}{P_t} \right)^2 \right]=0  \ .
\end{equation}

The dynamics is thus similar to that of a classical point particle of unit mass and zero total energy subject to the 
effective potential:
\begin{eqnarray}
\label{effective_potential}
  V_{\rm eff}(r)  &=&   \frac{1}{2}  \left[ -\left( \frac{r}{r_\sigma} \right)^{2 a} +
\frac{\epsilon}{P_t^2} \left( \frac{r}{r_\sigma} \right)^{4 a} \right.   \left. +
\left( \frac{P_z}{P_t} \right)^2 \left( \frac{r}{r_\sigma} \right)^{4 a - 2 b} + 
\frac{1}{\gamma r^2} \left( \frac{P_\theta}{P_t} \right)^2 \left( \frac{r}{r_\sigma} 
\right)^{4 a - 2 c} \right] 
  \end{eqnarray}
  
The equations for $\theta$ and $z$ are directly obtained from the constants of motion and read
\begin{equation}
\frac{dz}{dt}=-\frac{P_z}{P_t}\left( \frac{r}{r_\sigma} \right)^{2(a-b)} \ \ , \ \ 
\frac{d\theta}{dt}=-\frac{1}{\gamma r^2} \frac{P_\theta}{P_t}
\left( \frac{r}{r_\sigma} \right)^{2(a-c)}  \ .
\end{equation}

\subsection{The effective potential}

Some properties of the orbits can be deduced from the form of the effective potential. 
First note that $V_{\rm eff} < 0$ in order for (\ref{eq4}) to have solutions. 

In the following we will discuss the possible orbits in this space-time.
We will define bound, i.e. planetary orbits as orbits on which particles
move from a maximal finite radius $r_{\rm max}$ to a minimal radius $r_{\rm min} > 0$ and back again
in the $x$-$y$-plane. Note that these orbits can extend to infinity in the $z$-direction for $P_z\neq 0$. 
In contrast to that we have unbound orbits on which particles approach the string from
infinity, move to a minimal radius $r_{\rm min}$ and then return to infinity in the $x$-$y$-plane.
In both cases, it is also of interest to know whether $r_{\rm min}$ can become zero, i.e. whether the
particle can reach the string axis. In the following, we will call orbits
that end at $r_{\rm min}=0$ ``terminating orbits''. The possible orbits are
summarized in Table I.

\begin{table}[h]
\begin{center}
\begin{tabular}{|cccc|}\hline
type  & turning points 
& range of r & orbit \\ \hline\hline
A  & 0  & 
\begin{pspicture}(-2,-0.2)(3,0.2)
\psline[linewidth=0.5pt]{->}(-2,0)(3,0)

\psline[linewidth=1.2pt](-2,0)(3,0)
\end{pspicture} 
& terminating escape orbit \\  \hline
B  & 1  & 
\begin{pspicture}(-2,-0.2)(3,0.2)
\psline[linewidth=0.5pt]{->}(-2,0)(3,0)
\psline[linewidth=1.2pt]{-*}(-2,0)(0,0)
\end{pspicture} 
 & terminating orbit \\ \hline
C  & 2  & 
\begin{pspicture}(-2,-0.2)(3,0.2)
\psline[linewidth=0.5pt]{->}(-2,0)(3,0)
\psline[linewidth=1.2pt]{*-*}(-0.5,0)(1.5,0)
\end{pspicture} 
 & bound orbit \\ \hline
D & 1 & 
\begin{pspicture}(-2,-0.2)(3,0.2)
\psline[linewidth=0.5pt]{->}(-2,0)(3,0)
\psline[linewidth=1.2pt]{*-}(0.5,0)(3,0)
\end{pspicture} 
 & escape orbit \\ \hline\hline
\end{tabular}
\caption{Types of possible orbits
for test particles moving in the static Kasner space-time.}
\end{center}
\end{table}

 \begin{figure}[h]
\begin{center}

\subfigure[][effective potential $V_{\rm eff}(r)$ for $b^+$, $c^+$]{\label{pot1}
\includegraphics[width=8.5cm]{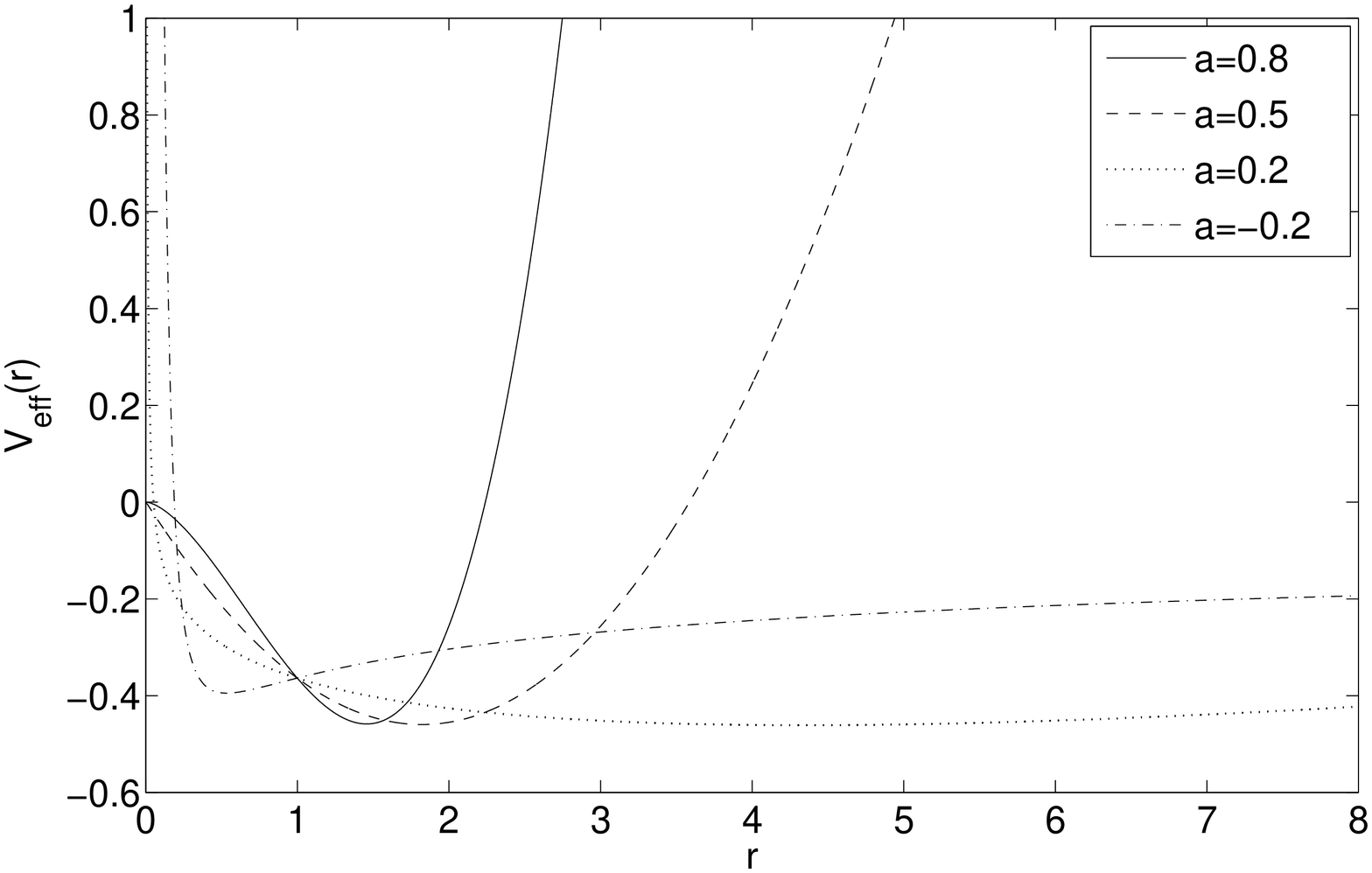}}
\subfigure[][effective potential $V_{\rm eff}(r)$ for $b^-$, $c^-$]{\label{pot2}
\includegraphics[width=8.5cm]{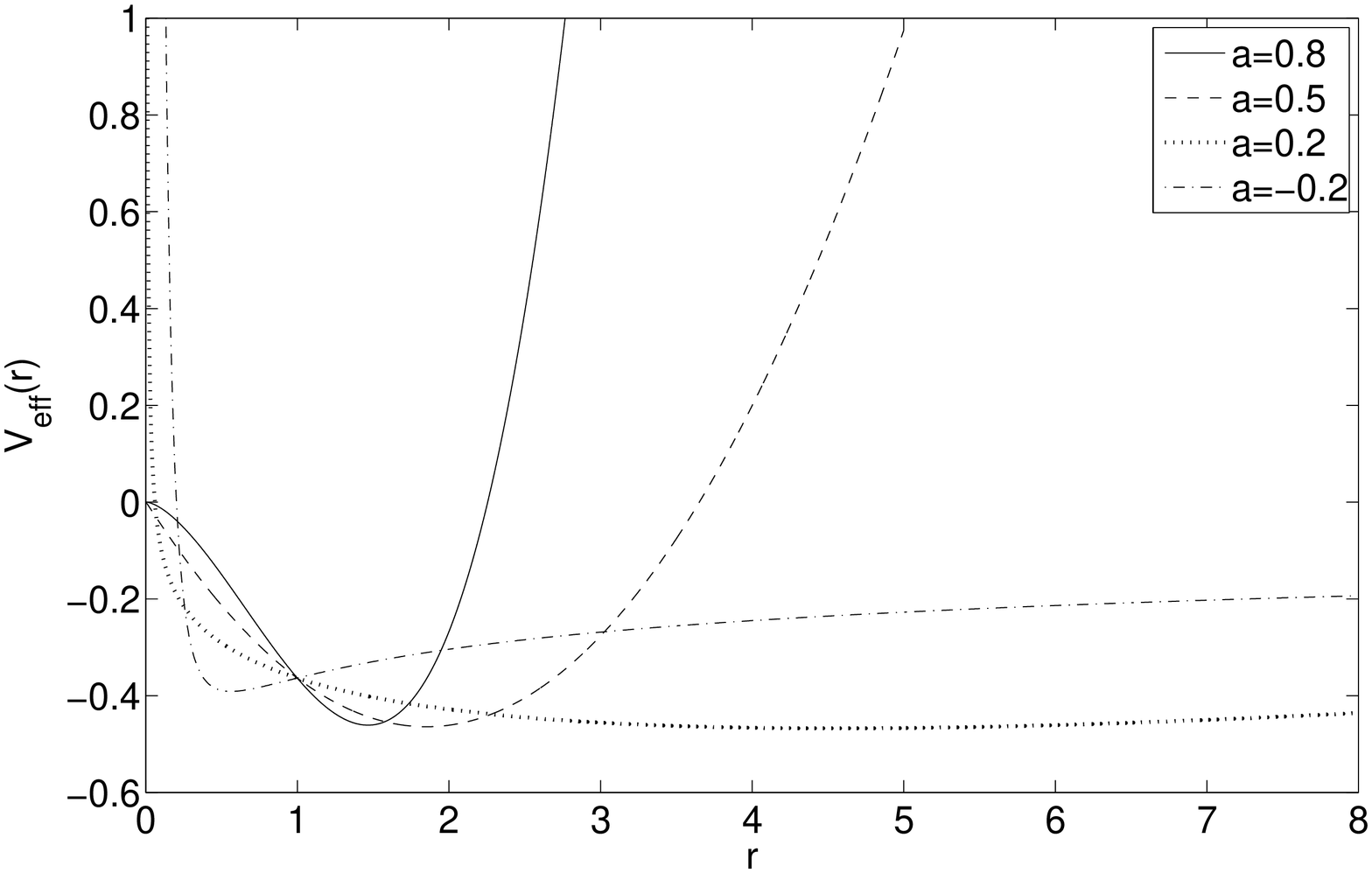}} 

\end{center}
\caption{\label{potentialV} We show the effective potential $V_{\rm eff}(r)$ for 
a massive test particle ($\epsilon=1$) with
$P_t=2$, $P_z=P_{\theta}=0.2$ in a Kasner space-time with $r_{\sigma}=1$, $\gamma=0.8$ and
different choices of $a$. Note that the Kasner conditions then fix the possible
values of $b$ and $c$. We show the effective potential for $b^+$, $c^+$ (left) as well
as for $b^-$, $c^-$ (right). }
\end{figure}

\begin{table}
\begin{center}
\begin{tabular}{|l|l|l|l|l|l|l|l|l|l|l|l|l|l|l|l|l|}
  \hline
  $a$ & \multicolumn{4}{|c|}{ $< 0$ } & \multicolumn{2}{|c|}{ $0$ } &
\multicolumn{6}{|c|}{$> 0$ and $<2/3$ } & \multicolumn{4}{|c|}{ $>2/3$ }
 \\
  \hline
  $\epsilon$ & \multicolumn{3}{|c|}{$0$ } & $-1$& \multicolumn{2}{|c|}{} &
\multicolumn{4}{|c|}{$0$ } &
\multicolumn{2}{|c|}{$-1$}&\multicolumn{3}{|c|}{ $0$ } & 1 \\ \hline
  $P_z$ & \multicolumn{2}{|c|}{$0$} & $\neq 0$ & & $0$ & $\neq 0$ &
\multicolumn{2}{|c|}{$0$} & \multicolumn{2}{|c|}{$\neq 0$} & $0$ & $\neq
0$ &\multicolumn{2}{|c|}{ $0$ } & $\neq 0$ &  \\ \hline
  $P_\theta$ & $0$ & $\neq 0$ & & & & & $0$ & $\neq 0$ & $0$ & $\neq 0$ &
& & \multicolumn{1}{|c|}{ $0$ } & $\neq 0$ & &  \\ \hline
  type & A & D & D & D & A & D & A & B & D & C & B & C & A &B &B &B \\ \hline
\end{tabular}
\end{center}
\caption{Types of orbits in the static Kasner space-time in dependence on
the different parameters for the $(b^+,c^+)$ solution. The table for
$(b^-,c^-)$ would be the same with $P_z$ and $P_\theta$ exchanged.}
\end{table}

\begin{figure}[h]
\begin{center}

\subfigure[][effective potential $V_{\rm eff}(r)$ for $b^+$, $c^+$]{\label{pot3}
\includegraphics[width=8.5cm]{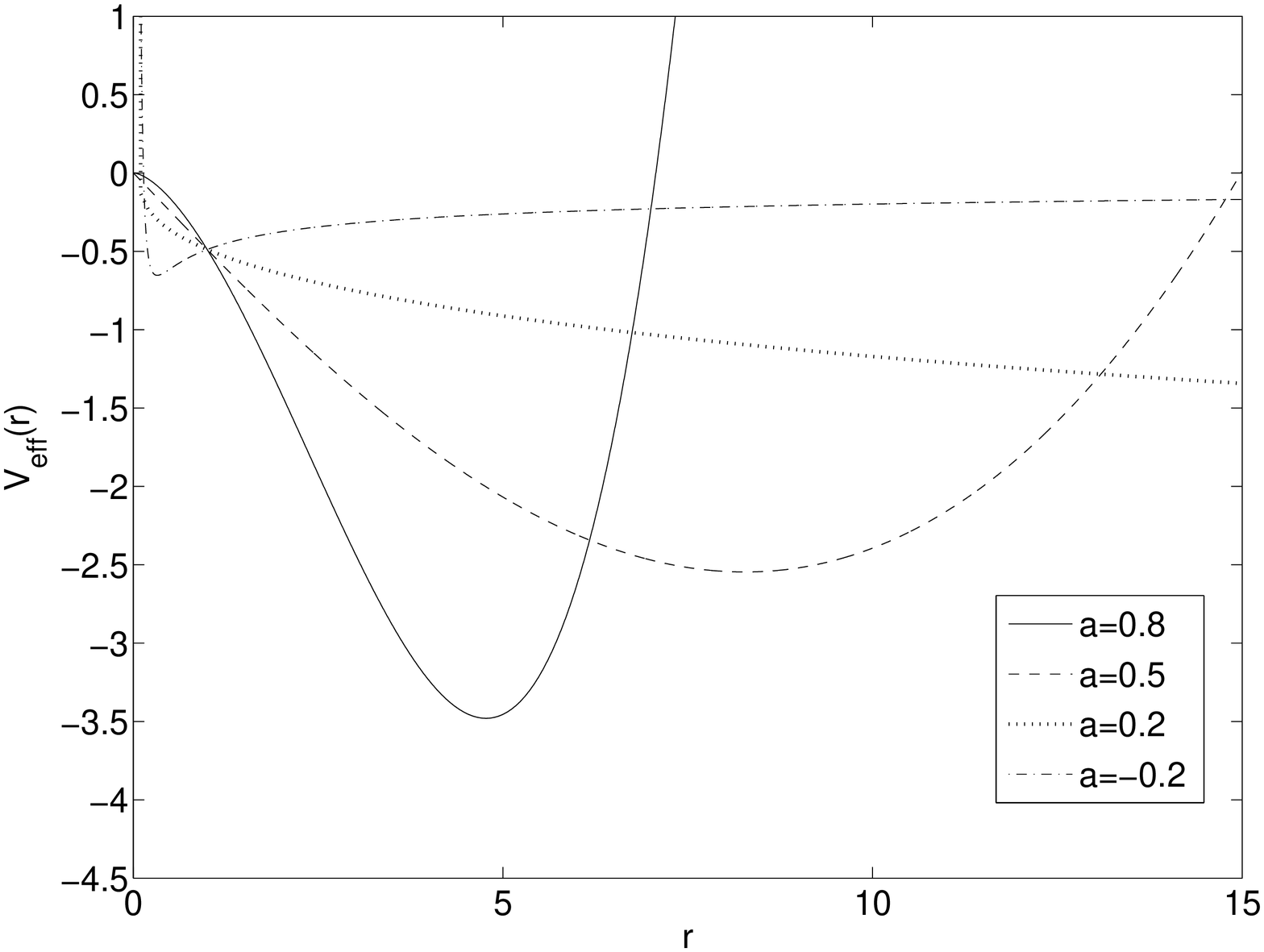}}
\subfigure[][effective potential $V_{\rm eff}(r)$ for $b^-$, $c^-$]{\label{pot4}
\includegraphics[width=8.5cm]{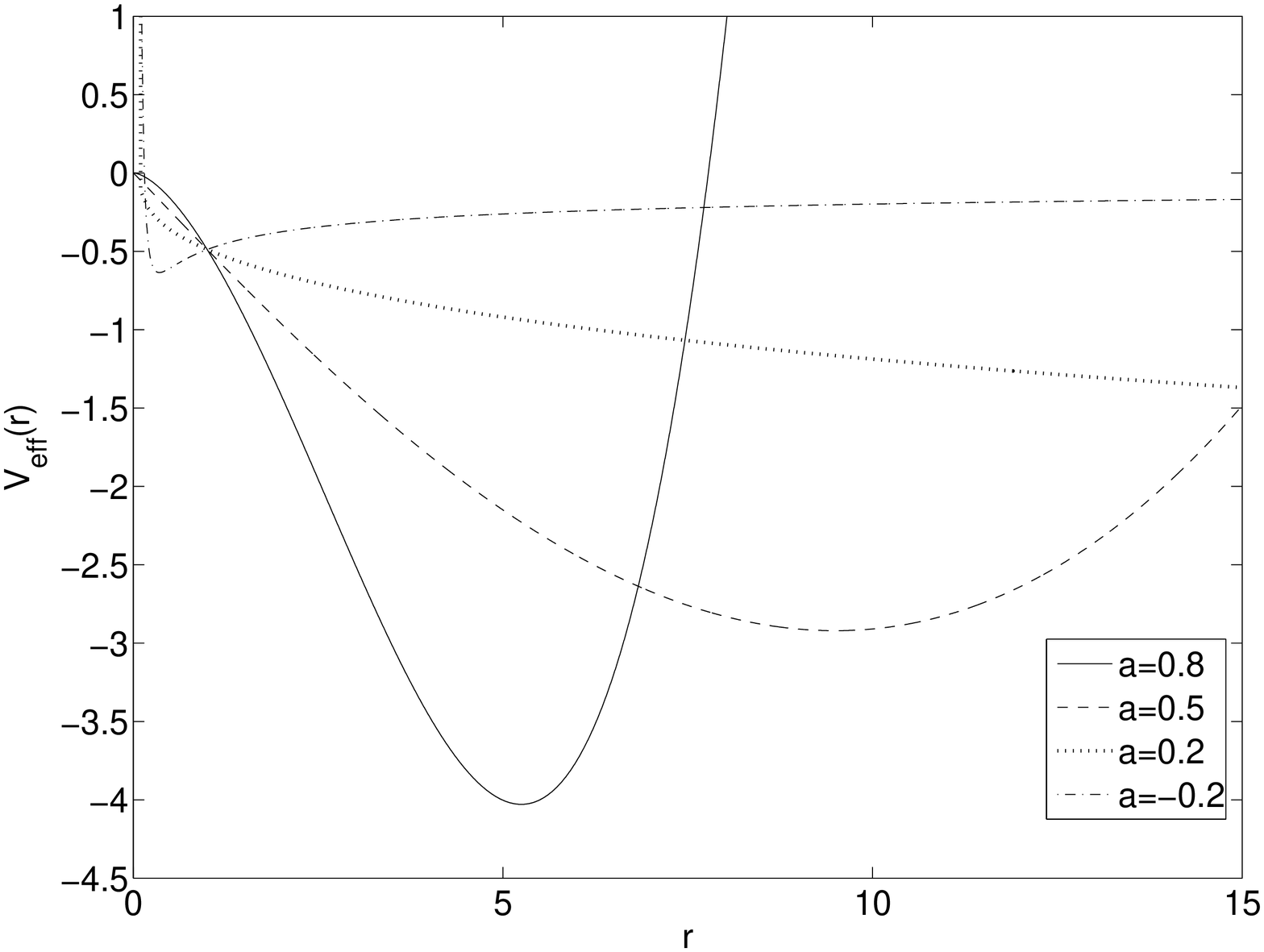}} 

\end{center}
\caption{\label{potentialVmassless} We show the effective potential $V_{\rm eff}(r)$ for 
a massless test particle ($\epsilon=0$) with
$P_t=2$, $P_z=P_{\theta}=0.2$ in a Kasner space-time with $r_{\sigma}=1$, $\gamma=0.8$ and
different choices of $a$. Note that the Kasner conditions then fix the possible
values of $b$ and $c$. We show the effective potential for $(b^+$, $c^+)$ (left) as well
as for $(b^-$, $c^-)$ (right). }
\end{figure}

In Fig.\ref{potentialV} and Fig.\ref{potentialVmassless} we show the effective
potential $V_{\rm eff}(r)$ for massive and massless particles, respectively, with
$P_t=2$, $P_z=P_{\theta}=0.2$ in the space-time of a Kasner metric with $r_{\sigma}=1$, $\gamma=0.8$ and
different choices of $a$. The choice of $a$ fixes the value of $b$ and $c$ (see (\ref{abc})). We give the 
potential for both choices of $b$ and $c$ and note that the difference between the two choices
is marginal. For general choices of $a$ our results are summarized in Table II. These are:

\begin{itemize}
 \item  If $a < 0$, there are no bound orbits since the potential has no local maximum and is negative at infinity.
This
is clearly seen in Fig.\ref{potentialV} and Fig.\ref{potentialVmassless} for $a=-0.2$. In addition,
the minimal radius $r_{\rm min}$ is finite, 
except for a massless particle with vanishing $P_z$ and $P_\theta$. 
In this case, $V_{\rm eff}(r) \propto r^{2 a}$ for $r \rightarrow 0$. Since 
$a <0 $, $\frac{dt}{dr} \propto \frac{1}{\sqrt{V_{\rm eff}}} \propto r^{-a}$ is integrable around 
$r=0$ and the particle thus reaches the string after a finite coordinate time.

\item For $a=0$, there are no bound orbits. This corresponds either to the case of a standard cosmic string in which
case the space-time is locally flat with $b=b^-=c^-=c=0$ or to $b=b^+=-c^+=-c=1$ which is not physical. 
The minimal radius $r_{\rm min}$ is generally finite, 
except if $P_z=0$ (for $b^+$, $c^+$) or $P_\theta=0$ (for $b^-$, $c^-$), in 
which case the potential is just a constant. 

\item  For $a > 0$ and $\epsilon = 1$ (massive particle): the orbits are always bound
if $a < 2/3$ and  $P_z \neq 0$ (for $b^+$, $c^+$) or $a < 2/3$ and 
$P_\theta \neq 0$ (for $b^-$, $c^-$). This is clearly seen in Fig.\ref{potentialV} for $a=0.2$ and $a=0.5$, where
the potential in both cases has positive values close to $r=0$ and tends to positive values at $r\rightarrow \infty$.
For other choices of $a$ the orbits are bound terminating, i.e. end at the string axis at $r=0$ since
the potential tends to zero from below at $r\rightarrow 0$ (see the potential
for $a=0.8$ in Fig.\ref{potentialV}).
When the minimal radius vanishes, 
the dominant term near $r=0$ is at least $r^{2 a}$. So, $\frac{dt}{dr}=\mathcal{O}(r^{-a})$. 
Since $a <1$, this is integrable around $r=0$ and the time needed to reach the string is 
finite (except if $a =1$). 
\item  For $a > 0$ and $\epsilon = 0$ (massless particle): the behaviour around 
$r=0$ is exactly the same as in the massive case (since the removed term is negligible),
 but the orbits are unbound for $a \leq 2/3$ when choosing $P_\theta=0$ 
  for ($b^+$, $c^+$)  or $P_z=0$ for ($b^-$, $c^-$). This is seen in Fig.\ref{potentialVmassless}.
This means that for appropriate choices of the parameters, we find that bound orbits for 
massless test particles exist. 
\end{itemize}

We remark that exchanging $P_\theta$ and $P_z$ does not change the qualitative form of the potential,
 provided we also exchange $(b^+,c^+)$ with $(b^-,c^-)$.
In fact, the equation of motion for $r$ is invariant under 
\begin{eqnarray}
(b^+,c^+) \leftrightarrow (b^-,c^-) \ \ , \ \ 
P_z \leftrightarrow \frac{P_\theta}{\sqrt{\gamma r_\sigma^2}}
\end{eqnarray}
so that we can restrict our attention to one of the two cases.

Studying the asymptotic properties of $V_{\rm eff}$ is \textit{a priori} not 
enough to rule out bound orbits, or to say that there is only 
one bound orbit (the potential could cross zero several times in the intermediate region.) 
However, we find that nothing new arises from a more detailed study.

Note that the case of interest in the context of cosmic strings with additional structure 
corresponds to $a > 0, a \approx 0, b= b^- \approx -a, c= c^- \approx 0$.   

\subsection{Numerical results}

The equations of motion are solved numerically using a second-order 
symplectic integrator, which avoids numerical dissipation effects. We studied the domain of existence
and the dependence 
of the maximal and minimal radius on the Kasner coefficients and constants of motion, 
as well as the light deflection using \textit{Maple} and \textit{Mathematica}.

\subsubsection{Domain of existence}

\begin{figure}[h]
\begin{center}

\subfigure[][$P_t=1$
  ]{\label{lz_pz_e1}
\includegraphics[width=8cm]{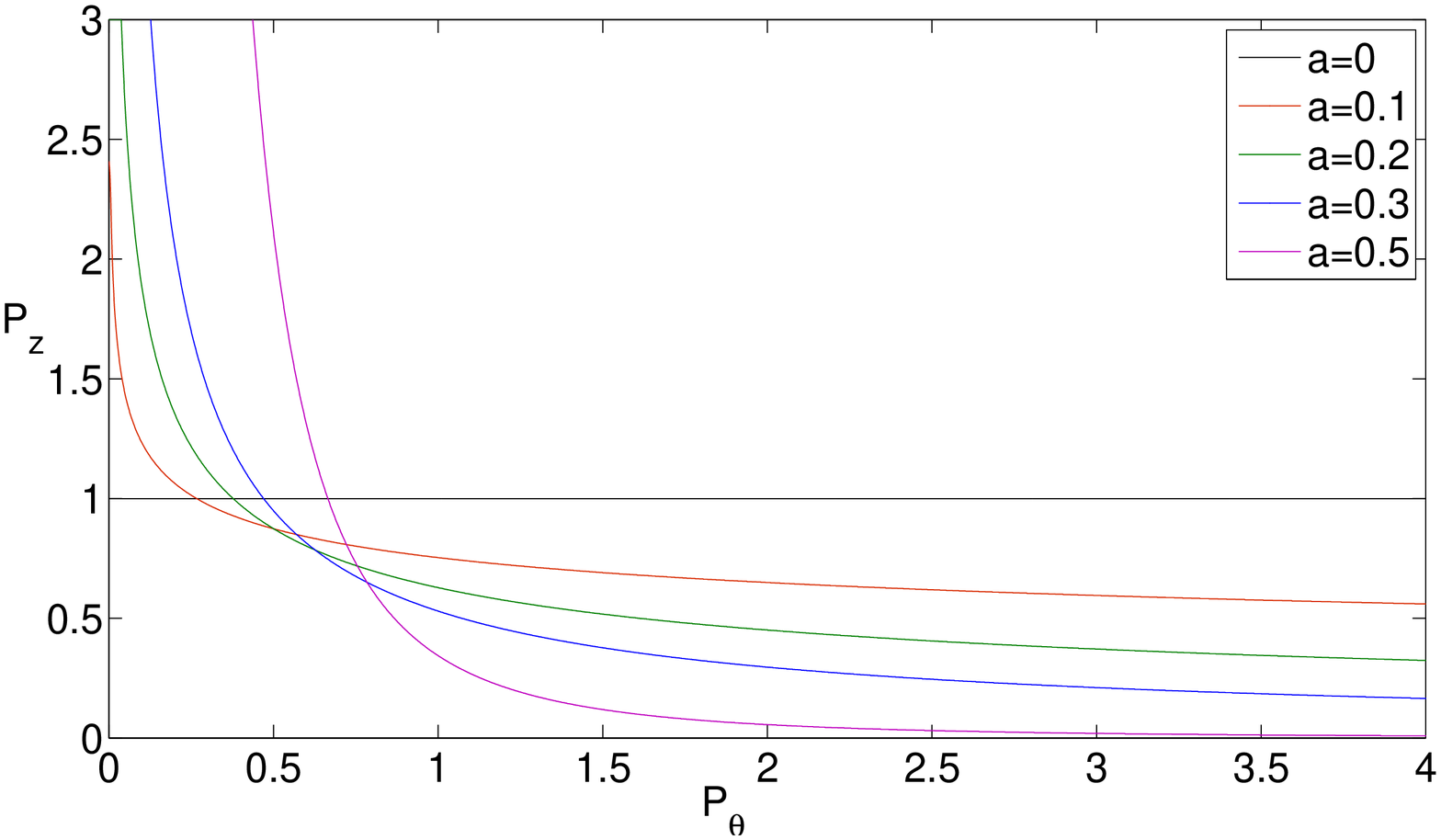}}
\subfigure[][$P_t=2$]{\label{lz_pz_e2}
\includegraphics[width=8cm]{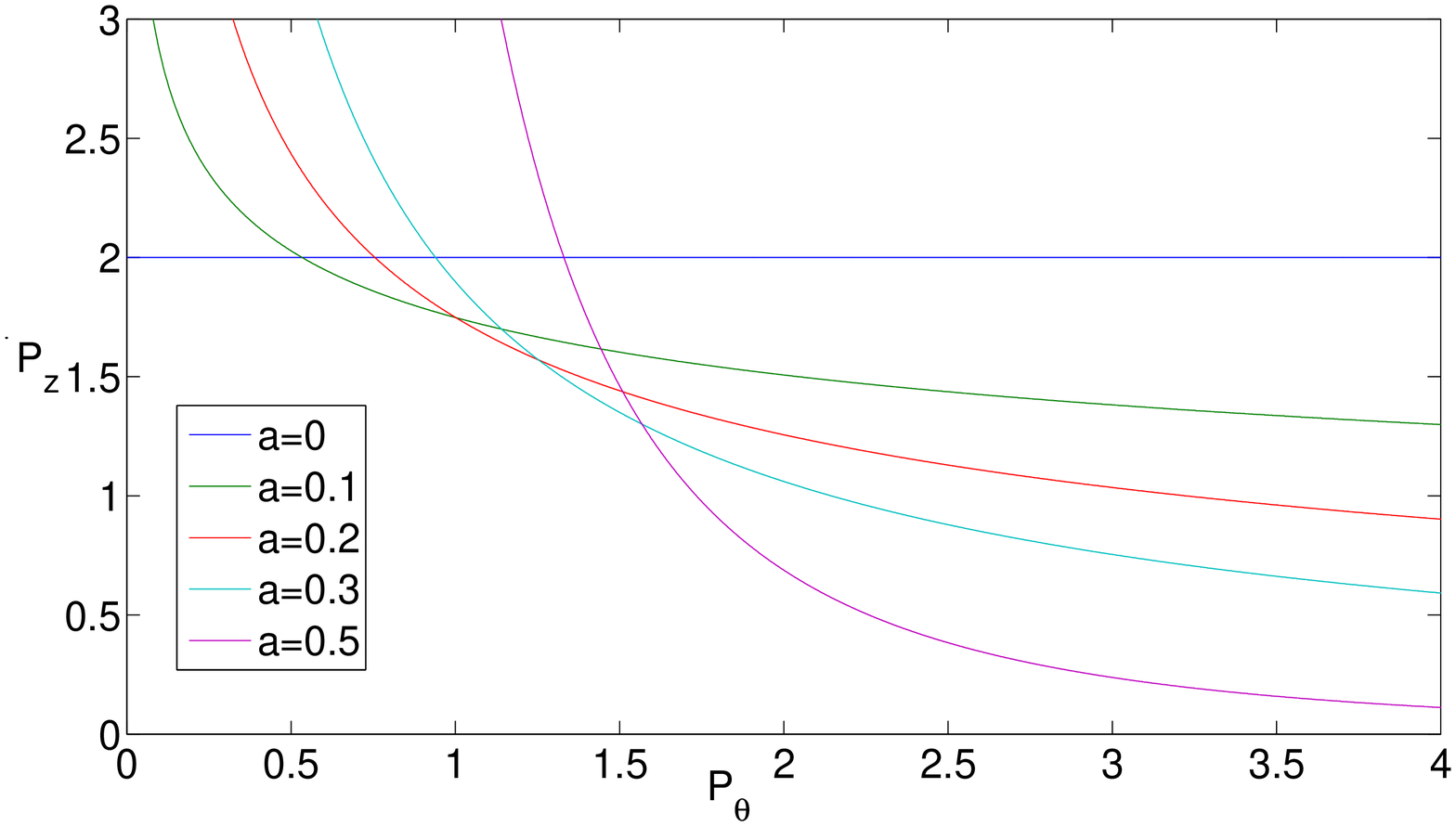}} 
\end{center}
\caption{\label{doe1} 
We show the domain of existence of solutions to the geodesic equation 
in the $P_{\theta}$-$P_z$-plane in a general
Kasner space-time with $r_{\sigma}=1$, $\gamma=1$ and different values of $a$ for massless test particles 
with $P_t=1$ (left) and $P_t=2$ (right). Note that solutions to the geodesic equation exist below and to the left
of the curves.
}
\end{figure}

Given the parameters $a$, $b$ and $c$ in the Kasner metric, a natural question is
what values of $\gamma$, $P_t$, $P_z$ and $P_\theta$ allow for solutions of the equations of motion. 
A necessary and sufficient condition is that the potential must be negative somewhere. 
For $a < 0$ or $a > 2/3$, the dominant term at zero or infinity respectively is 
$- \frac{1}{2} \left( \frac{r}{r_\sigma} \right)^{2 a}$. Therefore, 
solutions always exist. For $a \in \left] 0,2/3 \right[$, 
the domain in which solutions exist can be computed numerically. 
For massless particles, since $r_\sigma$ can be taken equal to one by rescaling $r$ 
and $\gamma$ can be taken to one by rescaling $P_\theta$, 
the function $r \mapsto V_{\rm eff}(r)$ depends only on two parameters, 
for instance on $P_z/P_t$ and $P_\theta/P_t$ for $r_\sigma=1$, $\gamma=1$. 
Moreover, because of the aforementioned duality, we can restrict 
our attention to the $(b^-,c^-)$ solution. For massive particles, 
$P_t$ becomes a physically relevant parameter.
The domain in which solutions exist tend towards the one obtained for $\epsilon=0$ in the limit $P_t \rightarrow \infty$ and shrink when $P_t$ decreases.
For $a=0$, we found analytically that solutions exist if and only if $P_z^2+\epsilon \leq P_t^2$ for the 
$(b^+,c^+)$ case or $\frac{P_\theta^2}{\gamma}+\epsilon \leq P_t^2$ for the $(b^-,c^-)$ case (in units $r_\sigma=1$).
For $a=2/3$, these conditions become $\left( \frac{P_z}{P_t} \right)^2 \leq 1$ or $\frac{1}{\gamma}\left( \frac{P_\theta}{P_t} \right)^2 \leq 1$ 
for the $(b^+,c^+)$ and $(b^-,c^-)$ case respectively.
Our results are summarized in Fig.\ref{doe1}  for massless particles and
particular choices of $P_t$. 

\subsubsection{Examples of orbits}
\begin{figure}[h]
\begin{center}

\subfigure[][massive test particle, $P_t=2$, $P_z=0$, $P_{\theta}=0.5$ ,
$a=0.3$, $b=-0.2266281297$, $c=-0.07337187027$ and $\gamma=0.5$   ]{\label{orbit_massive}
\includegraphics[width=7cm]{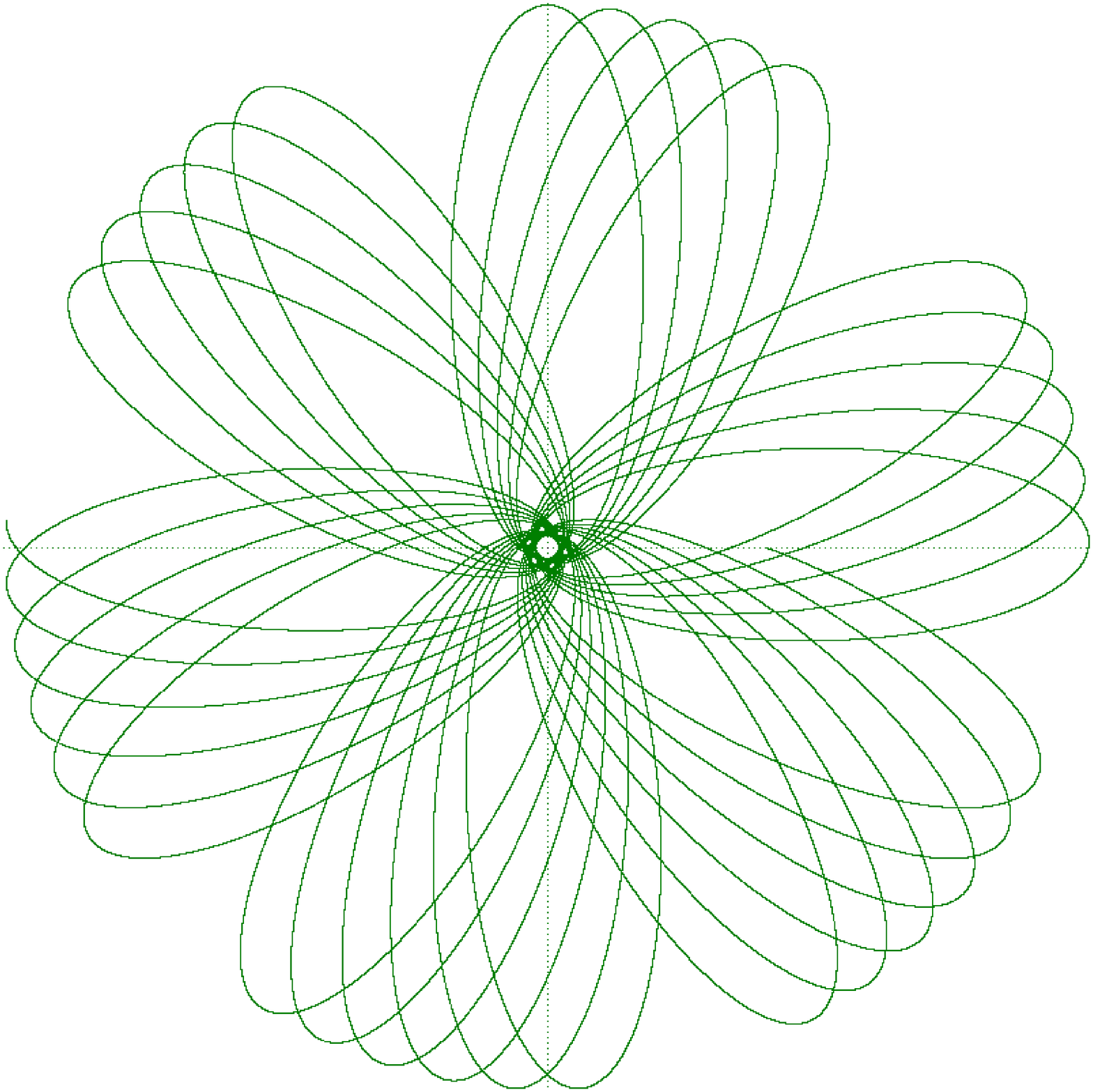}}
\subfigure[][massless test particle, $P_t=2.0$, $P_z=0.1$, $P_{\theta}=0.03$ ,
$a=0.8$, $b=0.512311$, $c=-1.31231$ and $\gamma=0.8$]{\label{orbit_massless}
\includegraphics[width=7cm]{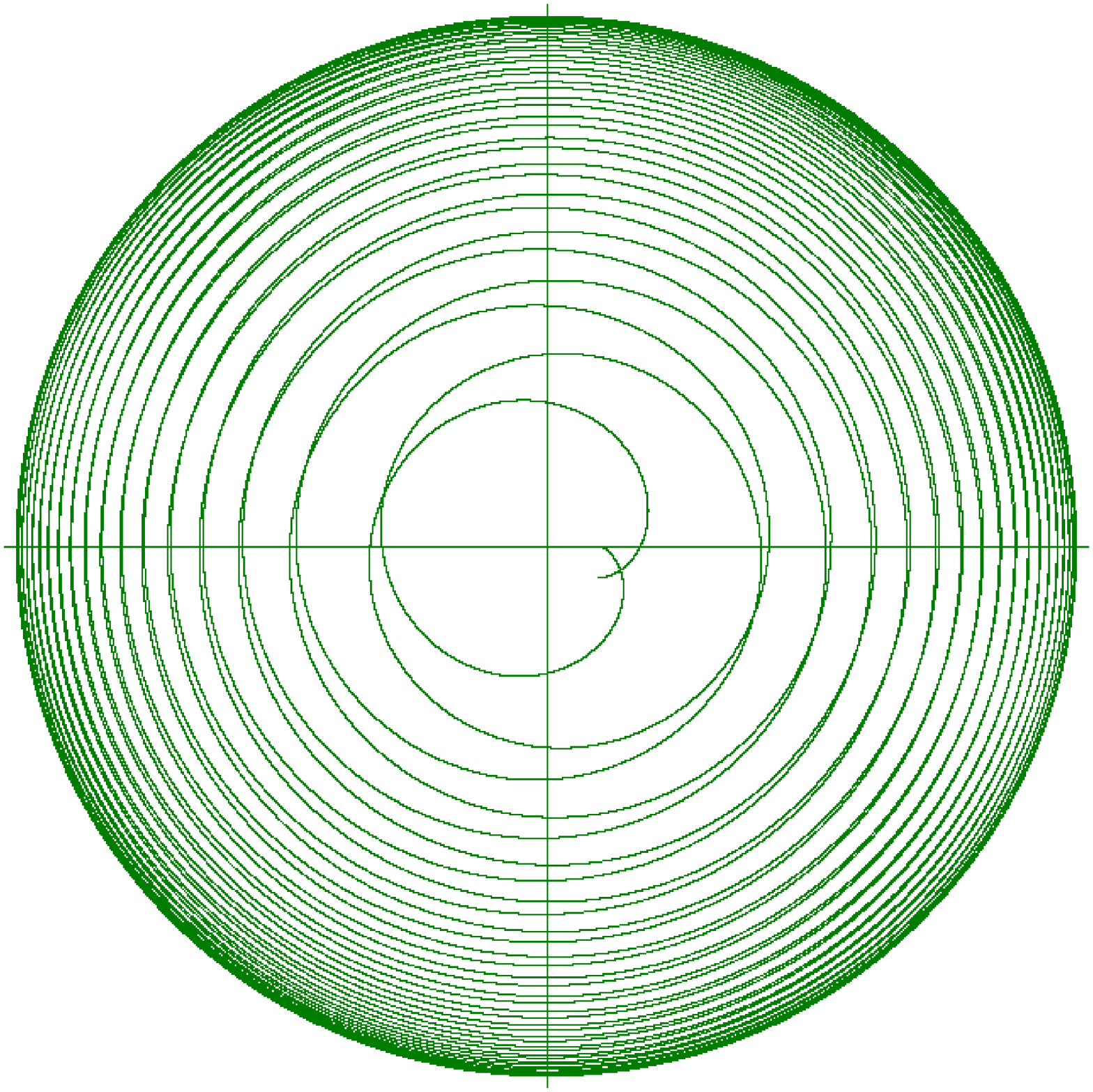}} 
\end{center}
\caption{\label{boundorbits1} 
We show a bound orbit for a massive (left) and a massless (right) test particle in a Kasner space-time.
}
\end{figure}

The most important difference to the case $T=U$ in which geodesics are just straight
lines because the space-time is locally flat is that we can have bound orbits for massive and massless
test particles in the
general Kasner space-time. In Fig.\ref{boundorbits1} 
we show bound orbits for a massive and massless test particle, respectively. The bound orbit of a massive test particle
shows the typical perihelion shift of orbits in curved space-time. The bound orbit of the massless test particle
also possesses this perihelion shift, however looks qualitatively rather like a spiral. We observe that this
in related to the choice of $a$, which is quite big. For $a$ small the orbit of a massless particle looks qualitatively similar
to that of a massive one.

\begin{figure}[h]
\begin{center}

\subfigure[][massive test particle]{\label{orbit_massive_escape}
\includegraphics[width=7cm]{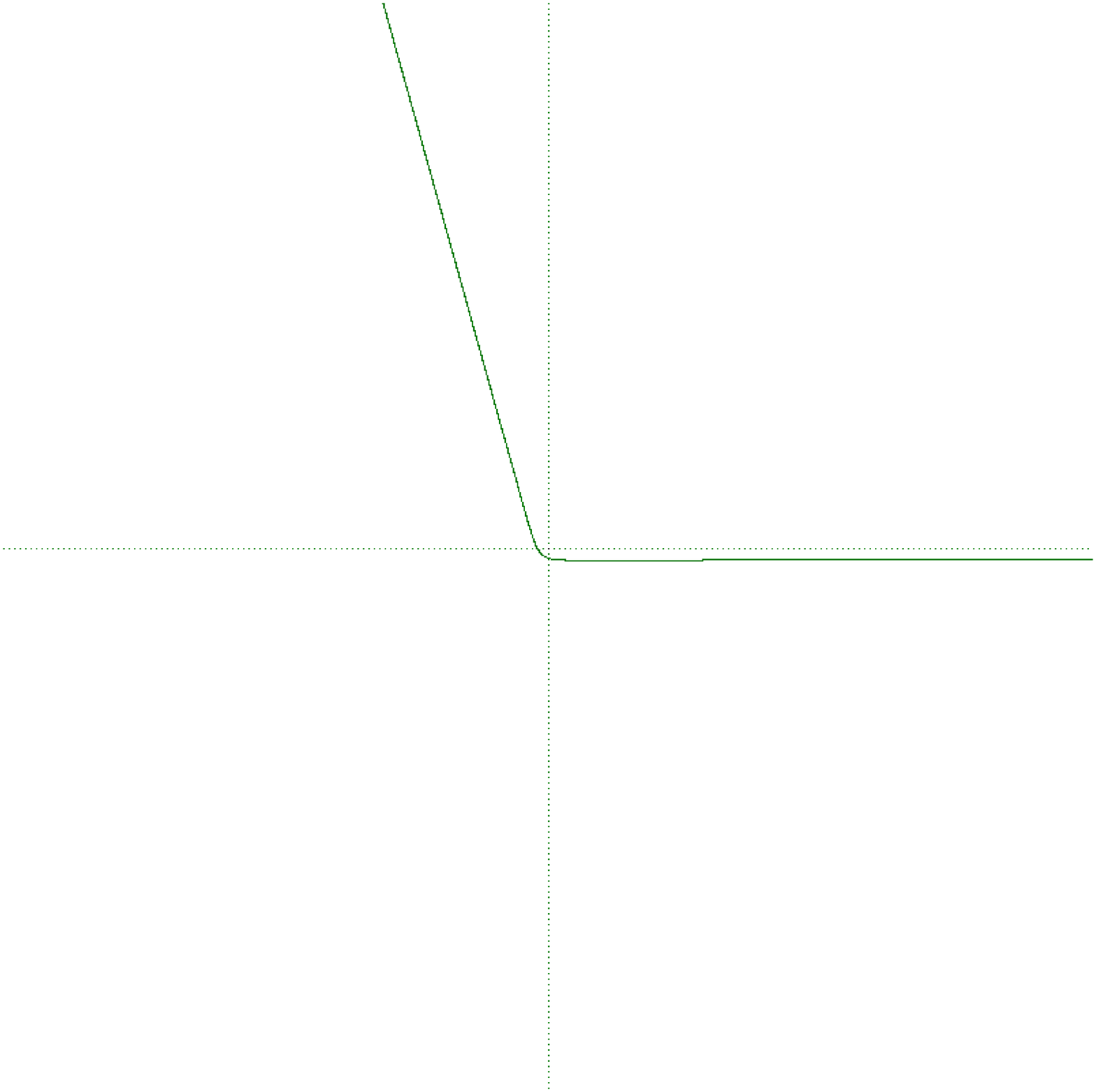}}
\subfigure[][massless test particle]{\label{orbit_massless_escape}
\includegraphics[width=7cm]{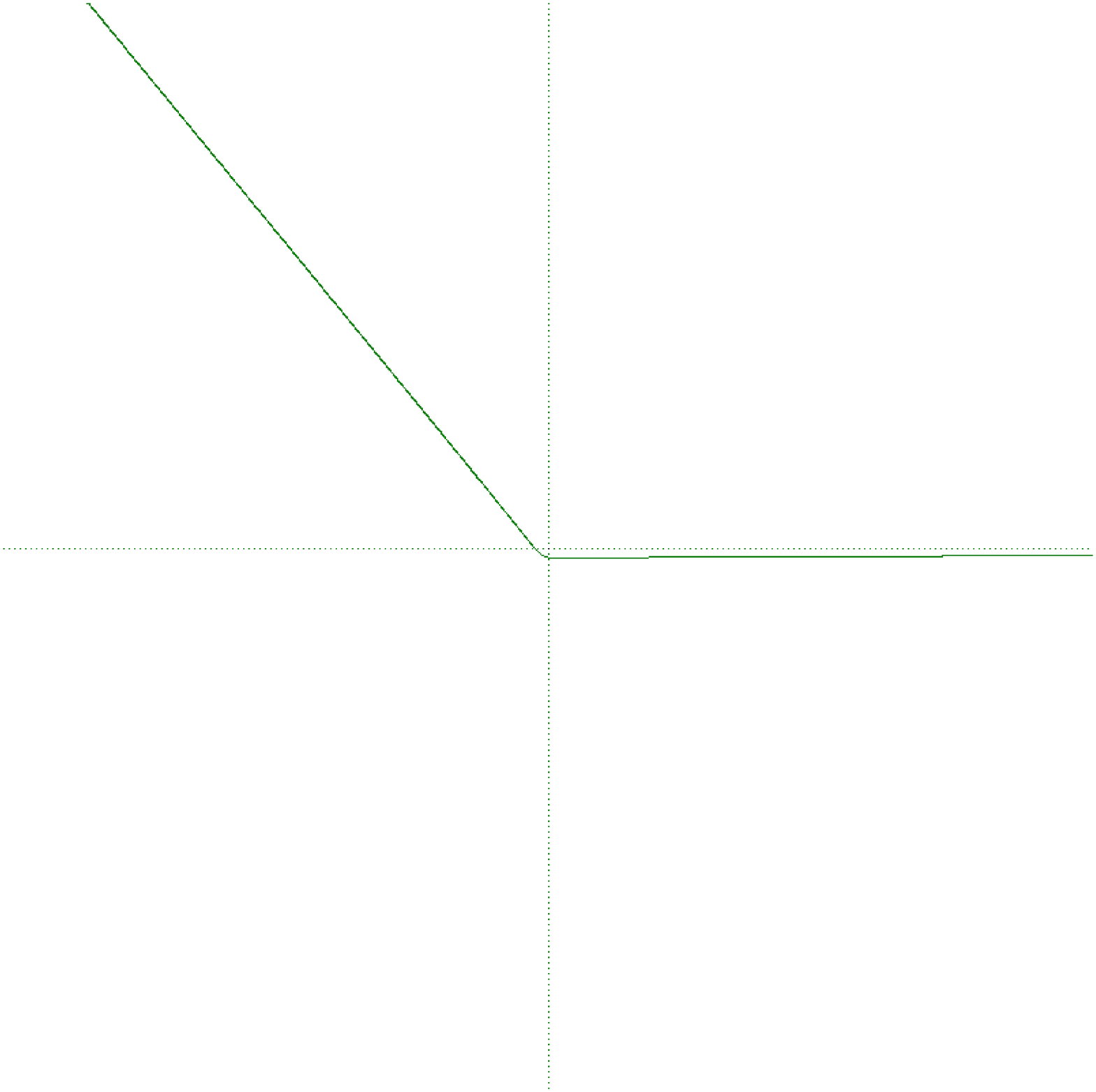}} 
\end{center}
\caption{\label{escapeorbits1} 
We show escape orbits for a massive test particle (left) and a massless test particle (right)
for  $P_t=2$, $P_z=0.1$, $P_{\theta}=0.2$ ,
$a=0$, $b=b_-$ and $c=c_-$ and $\gamma=0.5$ in a Kasner space-time.
}
\end{figure}

In Fig.\ref{escapeorbits1} 
we show escape orbits for a massive and massless test particle, respectively. In both cases, the test particles
get deflected by the string, which is related to the presence of the deficit angle
of the space-time. We observe that the massive test particle experiences a stronger deflection
for the same values of all parameters as the massless one.

\subsubsection{Minimal and maximal radius of bound orbits}

\begin{figure}
\centering
\epsfysize=8cm
\mbox{\epsffile{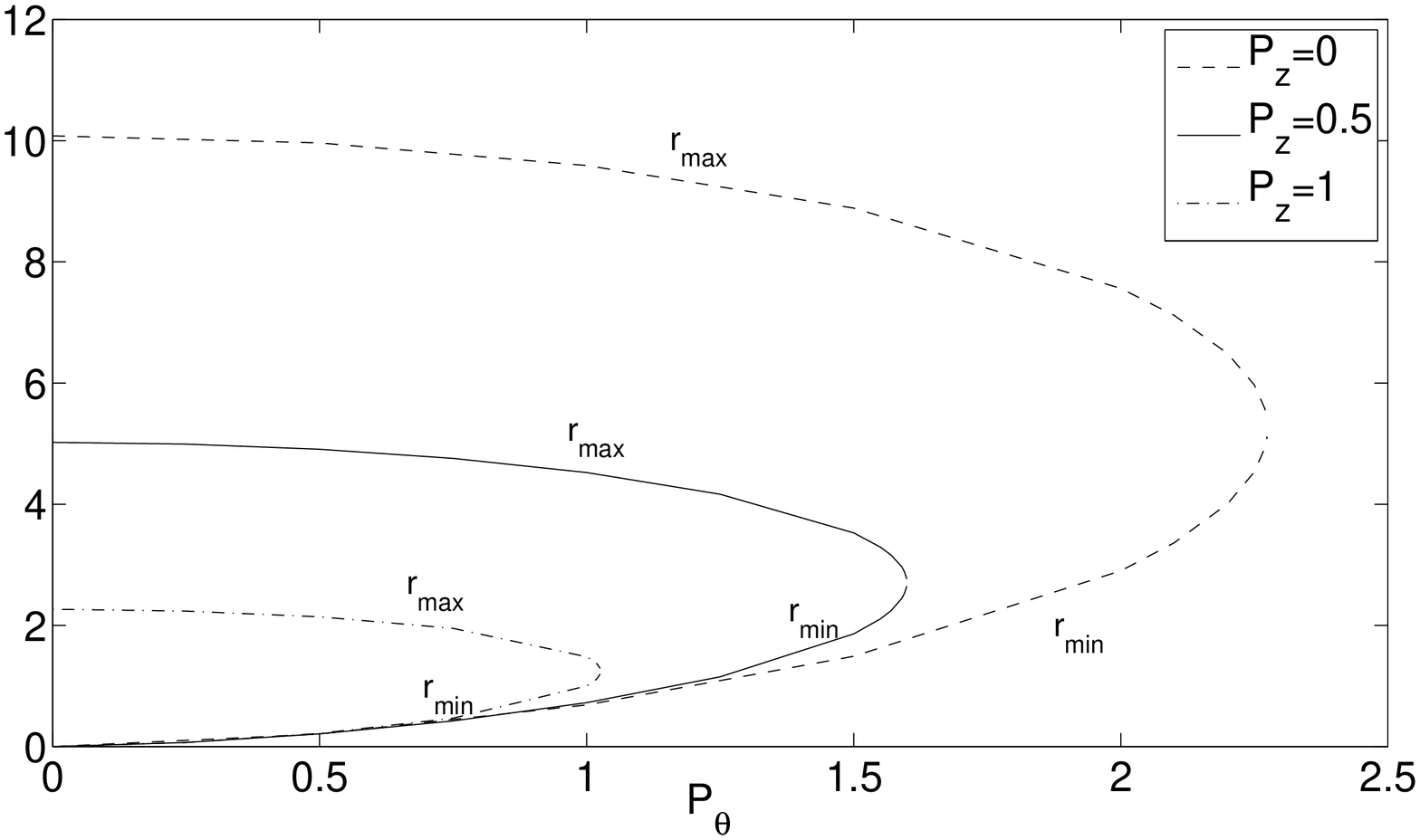}}
\caption{\label{rmin_rmax}
The value of the minimal radius $r_{\rm min}$ and of the maximal radius $r_{\rm max}$ of
bound orbits of massive test particles $(\epsilon=1$) with $P_t=2$ in a Kasner
space-time with $a=0.3$, $b^-$, $c^-$, $\gamma=0.5$ are shown in dependence of $P_{\theta}$ for three different values of $P_z$.}
\end{figure}

If the momenta are held fixed, the minimal and maximal radii of the bound orbits are decreasing functions of $a$. 
The maximal radius generally goes to infinity when $a \rightarrow 0^+$, 
while the minimum radius remains finite. This was to be expected since the orbits are 
unbound with finite minimal radius for $a < 0$ (let alone the very special case $P_\theta = P_z=0$).
For $a > 2/3$ and fixed $P_\theta$, $P_t$, we 
find terminating bound orbits. The maximal radius 
decreases with $P_z$, going to zero at infinity. It remains finite as $P_z \rightarrow 0$, 
which was also expected from the above analysis.
For $0< a < 2/3$, there are two critical values of $P_z$ 
above which the equations of motion can not be satisfied for $(b^+$, $c^+)$ or $(b^-$, $c^-)$.  
The minimal and maximal radius merge at this point, and move apart from each other when we decrease $P_z$. 
The merging of the minimal and maximal radius corresponds to a local minimum of the effective potential.
If we consider the $(b^+,c^+)$ case, the maximal radius remains finite 
when $P_z \rightarrow 0$, while the minimum one goes to zero. For the $(b^-,c^-)$ case, it is $r_{min}$ 
which remains finite while $r_{max}$ goes to infinity if $\epsilon=0$. 
For $a <0$, we find no maximal radius, and the minimal one increases with $P_z$.    
In Fig.\ref{rmin_rmax} we show the minimal
and maximal radius of the bound orbits for a Kasner space-time with
$a=0.3$, $b=-0.2266281297$, $c=-0.073371870270$ and $\gamma=0.5$ in dependence on $P_{\theta}$.
The test particle has $P_t=2$ and three different values of $P_z$. 
The $r_{\rm min}$ and $r_{\rm max}$
curves meet at the circular orbit. We find that the interval of $P_{\theta}$ 
for which bound orbits exist increases when decreasing $P_z$ for a fixed $P_t$.  
Moreover, we observe that the radius of the circular orbit decreases with increasing $P_z$.
In addition, the circular orbit appears at smaller angular momentum $P_{\theta}$ when increasing
$P_z$.

\subsubsection{Light deflection}

For an unbound orbit, we can define the deviation angle as:
\begin{equation}
\Delta \theta  = 2 \int_{r_{min }}^{\infty } \left|\frac{d\theta }{dr}\right| \, dr-\pi = 
\frac{\sqrt{2}}{\gamma }\frac{P_{\theta }}{P_t} \int _{r_{min}}^{\infty }\frac{1}{r^2}
\left(\frac{r}{r_{\sigma }}\right)^{2a -2c}\frac{1}{\sqrt{-V_{\rm eff}}}dr-\pi
\end{equation}
where $r_{min}$ is the minimal distance from the string to the particle. 

If all the other parameters are fixed, $\Delta \theta$ is a decreasing 
function of $\gamma$. It goes to infinity when $\gamma \rightarrow 0$ and is equal 
to $\left(\frac{1}{\sqrt{\gamma }}-1\right) \pi$ in conical space-time. 
(This is true in a coordinate system in which the metric has the Kasner form and $\theta$ 
goes from $0$ to $2 \pi$. A geometrical argument shows that it reduces to the usual 
$2 \pi (1- \sqrt{\gamma})$ in a locally flat coordinate system.)

For $a >0$, the $(b^-,c^-)$ case
gives an unbound orbit for massless particles with vanishing $P_z$. We 
find that the deviation angle increases with $a$.

The deviation angle also seems to behave logarithmically in $P_\theta$ in the 
limit $P_\theta \rightarrow \infty$. For $P_{\theta}\rightarrow 0$ we have $\Delta \theta \rightarrow -\pi$.
This is related to the repulsive effect of the string: if the angular momentum vanishes, 
the particle just goes back to where it comes from.

\section{Summary and discussions}
In this paper we have studied superconducting string solutions of the $U(1)_{\rm local} \times U(1)_{\rm global}$ model
in curved space-time. For small values of the ratio between the symmetry breaking scale and the Planck mass
we find that the metric outside the string core matches the Kasner metric very well and the Kasner
coefficients can be given in terms of the energy per unit length $U$ and the tension $T$ as originally suggested
in \cite{patrick1}. In order to be able to decide about the macroscopic stability of these
objects using the stability conditions given by Carter \cite{carter,carter2} we have to 
know the metric functions on the full interval $r\in[0:\infty[$. We have hence integrated the full coupled
system of differential equations numerically and computed the energy per unit length and tension.
We find that the coupling to gravity can stabilize the strings with large charge number density for sufficiently large values
of the ratio between the symmetry breaking scale and the Planck mass and that in general 
the phase frequency threshold is absent in curved space-time. As such, the energy per unit length and tension
never diverge, but tend to more or less constant values for larger values of the charge number density.

We have also studied the motion of test particles in the general Kasner space-time and find that
the fact that the energy per unit length is non-equal to the tension can lead to bound orbits of
massive and massless test particle. The question is then whether these orbits are of interest for astrophysical
or cosmological applications. We find that the radius of these orbits is too large to be
of interest and that e.g. the gravitational wave emission from a particle moving on a bound orbit
around a cosmic string would be far too small to be detectable. This -- in turn -- means
that the assumption of an infinitely thin cosmic string that is often used in simulations of string networks
is a valid assumption and that the additional structure on the string does not have a big influence on the results.
\\
\\
\\
\\

{\bf Acknowledgments} 
We would like to thank Patrick Peter for many fruitful and enlightening discussions
as well as the Institut d'Astrophysique de Paris (IAP) where part of this work was carried out for its hospitality.
This work was partially funded by Deutsche Forschungsgemeinschaft (DFG) grant HA-4426/5-1. 
BH also gratefully acknowledges support within the framework of the DFG Research
Training Group 1620 {\it Models of gravity}. FM would like to thank 
the \'Ecole Normale Sup\'erieure, Paris, France.

\end{document}